\documentclass[fleqn,usenatbib]{mnras}

\usepackage{newtxtext,newtxmath}

\usepackage[T1]{fontenc}
\usepackage{ae,aecompl}

\usepackage{graphicx}	
\usepackage{amsmath}	
\usepackage{subfig}
\usepackage{pifont}
\usepackage[table]{xcolor} 

\def\mpcoh{\,h^{-1}{\rm cMpc}}

\def\citejap#1{\citeauthor{#1}\ \citeyear{#1}}

\providecommand{\e}[1]{\ensuremath{\times 10^{#1}}}

\def\JAP#1{#1}

\def\m@th{\mathsurround=0pt }
\def\eqalign#1{\null\,\vcenter{\openup1\jot \m@th
 \ialign{\strut\hfil$\displaystyle{##}$&$\displaystyle{{}##}$\hfil
 \crcr#1\crcr}}\,}

\title{The fate of baryons in counterfactual universes}

\author[Boon Kiat Oh et al.]{%
Boon Kiat Oh,$^{1, 2}$\thanks{E-mail: bkoh@roe.ac.uk}
John A. Peacock,$^{1}$
Sadegh Khochfar,$^{1}$
and Britton D. Smith$^{1}$
\\
$^{1}$Institute for Astronomy, University of Edinburgh, Royal Observatory, Edinburgh EH9 3HJ, United Kingdom\\
$^{2}$Center for Theoretical Physics, Department of Physics and Astronomy, Seoul National University, Seoul 08826, Korea\\
}

\pubyear{2021}

\begin{document}
\label{firstpage}
\pagerange{\pageref{firstpage}--\pageref{lastpage}}
\maketitle

\begin{abstract}
	We present results from nine simulations that compare the standard $\Lambda$ Cold Dark Matter cosmology ($\Lambda$CDM) with counterfactual universes, for approximately $100\,{\rm Gyr}$ using the {\tt Enzo} simulation code. We vary the value of $\Lambda$ and the fluctuation amplitude to explore the effect on the evolution of the halo mass function (HMF), the intergalactic medium (IGM) and the star formation history (SFH).  The distinct peak in star formation rate density (SFRD) and its subsequent decline are both affected by the interplay between gravitational attraction and the accelerating effects of $\Lambda$. 
	The IGM cools down more rapidly in models with a larger $\Lambda$ and also with a lower $\sigma_8$, reflecting the reduced SFRD associated with these changes -- although changing $\sigma_8$ is not degenerate with changing $\Lambda$, either regarding the thermal history of the IGM or the SFH. However, these induced changes to the IGM or ionizing background have little impact on the calculated SFRD. We provide fits for the evolution of the SFRD in these different universes, which we integrate over time to derive an asymptotic star formation efficiency. Together with Weinberg's uniform prior on $\Lambda$, the estimated probability of observers experiencing a value of $\Lambda$ no greater than the observed value is 13\%, substantially larger than some alternative estimates. Within the {\tt Enzo} model framework, then, observer selection within a multiverse is able to account statistically for the small value of the cosmological constant, although $\Lambda$ in our universe does appear to be at the low end of the predicted range.
\end{abstract}




\section{Introduction}

In the standard $\Lambda$CDM cosmological model, structure formation  happens bottom-up: early sub-galactic clumps of dark matter undergo hierarchical merging to create the spectrum of dark-matter haloes, which act as the arena for galaxy formation in the later universe
\citep{1982ApJ...263L...1P, 1988ApJ...327..507F, 1991ApJ...379..440B, 1993MNRAS.262..627L}.
Baryons, in the form of gas, fall into the potential wells defined by the dark matter haloes. The gas undergoes radiative cooling and can collapse to form stars, which immediately begin to return feedback energy into the surrounding medium. This process is self-regulatory: increased feedback is associated with a higher star formation rate, which inhibits further star formation and vice versa \citep{1991ApJ...379...52W, 2000MNRAS.319..168C,2003ApJ...599...38B}.

Proving the correctness of this picture is less straightforward than might have been hoped.
There is ample evidence that the framework of dark matter evolves in a manner close to prediction, both in its overall amplitude (\citejap{2017MNRAS.465.1454H}) and in the form of the halo mass function (\citejap{2017A&A...608A..65B}; \citejap{driver2022}), but the paradoxical outcome of feedback is that galaxy assembly is anti-hierarchical, with galaxies of higher stellar mass completing their assembly before smaller and more fragile systems (\citejap{2008ApJ...675..234P}). In the face of this complexity, it has been common to follow \citet{1996ApJ...460L...1L} in taking a global approach that focuses on the total star formation rate density (SFRD) as contributed by the whole galaxy population.
\citet{2014ARA&A..52..415M} compiled observational data from a multitude of infrared (IR) and ultra-violet (UV) surveys to derive an analytical fit to the cosmic SFRD since $z\approx8$: this peaks around $z=2$ and then declines by an order of magnitude by $z=0$. 

This shutdown or `quenching' of cosmic star formation arises from the complicated interplay between the formation and evolution of dark matter haloes and the self-regulation of baryonic processes, but it is interesting to note that it occurs just as the universe enters the phase of exponential expansion driven by `dark energy'. Although one of the goals of modern cosmology is to determine whether the dark energy itself undergoes any evolution, the simplest hypothesis is that we are dealing with a cosmological constant, $\Lambda$, and we assume this hereafter.
Despite its importance, $\Lambda$ was never a permanent component of the cosmological model for many decades following its invention in 1917 by Einstein, based on his assumption of a static universe.\footnote{This paper can be found in English translation at \\ {\tt https://einsteinpapers.press.princeton.edu/vol6-trans/433}}
But during the 1990s, there was an accumulation of evidence in favour of $\Lambda$ as a necessary ingredient of the cosmological model, starting with arguments from LSS+CMB by \citet{1990Natur.348..705E}, and confirmed with particular directness by the well-known observations of type Ia SNe \citep{1998AJ....116.1009R,1999ApJ...517..565P}. Despite occasional challenges, we have thus lived for over two decades with the consensus view that the expansion of the universe accelerates, driven by the repulsive gravitational properties of a small positive cosmological constant.

However, the value of $\Lambda$ is problematic from the point of view of fundamental physics. Any geometrical cosmological constant on the LHS of Einstein's equations can be taken to the RHS, where it combines with the physical vacuum density into a single effective value of $\Lambda$. But the vacuum contributions are expected to be hugely larger than the measured effective value, requiring an unexplained cancellation with the bare cosmological constant to implausible precision. The simplest estimate of the vacuum density arises from summing zero-point energy over electromagnetic wave modes up to some cutoff energy \citep{1967JETPL...6..316Z, 1968SvPhU..11..381Z,1989RvMP...61....1W}; as is well known, a cutoff at the Planck energy gives an effective $\Lambda$ around $10^{120}$ times the observed value, and even reducing the scale of `new physics' to 10\,TeV still leaves a discrepancy of a factor $10^{66}$. However, this common calculation is seriously flawed, as it is nonrelativistic and fails to yield a correct equation of state for the vacuum density. Since the argument is a minor modification of the calculation for black-body radiation, it is clear that the predicted equation of state is
$w\equiv P/\rho c^2 = 1/3$, rather than $w=-1$. A relativistic calculation yields a different result: $\rho_{\rm vac}\sim m^4\ln(m/M)$ in natural units, where $m$ is the mass of the particle for the field under study, and $M$ is the cutoff scale \citep{2011arXiv1105.6296K,2012CRPhy..13..566M}. This is very different to the naive $M^4$, and it vanishes for a massless field like electromagnetism. But in the end, because particles with $m\sim 1$\,TeV exist, the fine tuning of the bare $\Lambda$ still needs to be carried out at a precision of one part in $10^{62}$. Even if one takes the unjustified step of asserting that somehow the vacuum density does not gravitate, there are still higher-order contributions from the gravitational interactions of virtual particles, as pointed out by \citet{1967JETPL...6..316Z}. These give a density $\sim(M/M_{\rm P})^2 M^4$, where $M_{\rm P}$ is the Planck mass, which is still too large by a factor $10^{36}$ if we take a cutoff at 10\,TeV.

These problems motivate the interest in dynamical dark energy, in which a vacuum density is presumed to be evolving dynamically from a natural large value towards zero. But it is difficult to escape the vacuum energy problem, as any dynamical Lagrangian for dark energy can have a constant added to it without changing the non-gravitational dynamics, and therefore we still need to understand why any such additive constant is so very small. Because of this fundamental puzzle, there has been a long-standing interest in an ensemble approach, in which the cosmological constant is considered to be a random variable, and where the observed value is subject to observer selection -- with large values not being observed because they would have the effect of suppressing cosmic structure formation. 
A universe with a large positive $\Lambda$ will experience an expansion so rapid that structure formation will freeze out and cease to grow.
For a recollapsing universe with large negative $\Lambda$, structures form late and close to the end of the universe's lifetime. In either case, although for different reasons, there will be a lack of galaxies and hence of observers in such universes. This probabilistic argument was made by \citet{1987PhRvL..59.2607W,1989RvMP...61....1W}, and refined by e.g. \citet{1995MNRAS.274L..73E}, \citet{Garriga1999}, \citet{2000astro.ph..5265W} and \citet{2007MNRAS.379.1067P}. Furthermore, \citet{1987PhRvL..59.2607W, 1989RvMP...61....1W} extended the argument to predict that the cosmological constant would be observed to be non-zero -- because observer selection disfavours large values but does not require $\Lambda$ to vanish exactly.

Under the heading of the `anthropic principle' \citep{1974IAUS...63..291C}, such arguments have been controversial, although there can be no objection to the weak form in which the presence of observers biases the observed properties of the universe (so that, for example, it is no surprise that we live at a time when the observed CMB temperature is $<1000$\,K). But the application to $\Lambda$ has the more radical need for a `multiverse' ensemble of distinct universes, within which the values of fundamental physical parameters are different. Such a concept was only implicit in Weinberg's original argument, but the idea has been given some subsequent support by developments elsewhere. An explicit multiverse ensemble is suggested by many models of inflation, where inflation ends at different times within multiple causally disconnected bubbles \citep{1983PhRvD..27.2848V, 1986MPLA....1...81L}. Furthermore, the lack of a unique prediction for the low-energy vacuum state in string theory has led to the `landscape' picture in which there can be an enormous number ($\sim 10^{500}$) of possible values for the effective cosmological constant. \citep{2003dmci.confE..26S}. In combination, inflation and the landscape thus allow an explicit scenario within which Weinberg's vision of observer selection for $\Lambda$ can be realised.

It has to be admitted that there is significant opposition to the anthropic approach, with many seeing it as an abandonment of attempts to calculate $\Lambda$ from first principles, and also questioning whether the multiverse hypothesis constitutes testable science. There are also difficulties in calculating the prior probability distribution of physical constants such as $\Lambda$ that are allowed to vary within the multiverse: the `measure problem' \citep{2011CQGra..28t4007F}. It is not our intention here to add to this debate, but instead to focus on the more astrophysical side of the problem. The above discussion shows that there is ample reason to be interested in counterfactual universes in which the cosmological parameters, particularly $\Lambda$, differ from the observed ones. Even if there is no multiverse, it is still interesting to understand the impact of cosmic acceleration on the build-up of the galaxy population. The early anthropic literature approached this question rather simply, by considering just the abundance of dark matter haloes that have the mass of a typical galaxy, and the immediate challenge is to model this process in more realistic detail. 

In this paper, we therefore present cosmological hydrodynamic box simulations of counterfactual universes, with the principal specific aim of studying the resulting cosmic star-formation history in galaxies as a function of $\Lambda$. This question has previously been addressed with the EAGLE numerical code by \citet{2018MNRAS.477.3727B}, and there have also been a number of semianalytic studies of the same problem \citep{Bousso1, Bousso2, 2017MNRAS.464.1563S, 2022arXiv220407509S}. But the importance of the issue justifies a diversity of approaches; as we will see, there are significant differences between our results and this existing work. 

We vary the value of the cosmological constant, $\Lambda$, within the $\Lambda {\rm CDM}$ model; we assume that such variations maintain a flat universe, as expected in an inflationary multiverse. We also consider modifications of the value of the fluctuation amplitude, $\sigma_8$, because raising $\Lambda$ reduces the final post-freezeout inhomogeneity, as does lowering $\sigma_8$. By comparing the two effects, we can diagnose the importance of when freezeout happens, in addition to the level of inhomogeneity at freezeout.  In all these computations, we need to allow for the effects of the UV background. The level of this radiation is determined by the star-formation history, but the properties of the IGM are in turn affected by the UV background, potentially changing the rate of star formation. We develop an iterative scheme that allows such effects to be treated self-consistently.

This paper is structured as follows. Section \ref{sec:ic-scaling} describes the method used to scale counterfactual cosmological parameters from the $\Lambda {\rm CDM}$ model. These will be inputs for the generation of the initial conditions. Section \ref{sec:cosmo-setup} discusses the code and setup used to evolve and post-process the simulation results. Section \ref{sec:evol-hmf-igm} presents the resulting evolution of
the halo mass function and the intergalactic medium.
Section \ref{sec:evol-sfr} then discusses the long-term history of the cosmic star-formation rate density (SFRD), and in particular its sensitivity to the UV background applied in each simulation. We then look in more detail at the cosmology dependence of the asymptotic star-formation efficiency in Section \ref{sec:asymptotic}  Lastly, Section \ref{sec:new-cosmo-conclusion} summarises the results and compares our conclusions with those of other authors who have considered this problem.

\begin{figure}
	\centering
	\includegraphics[width=.98\linewidth]{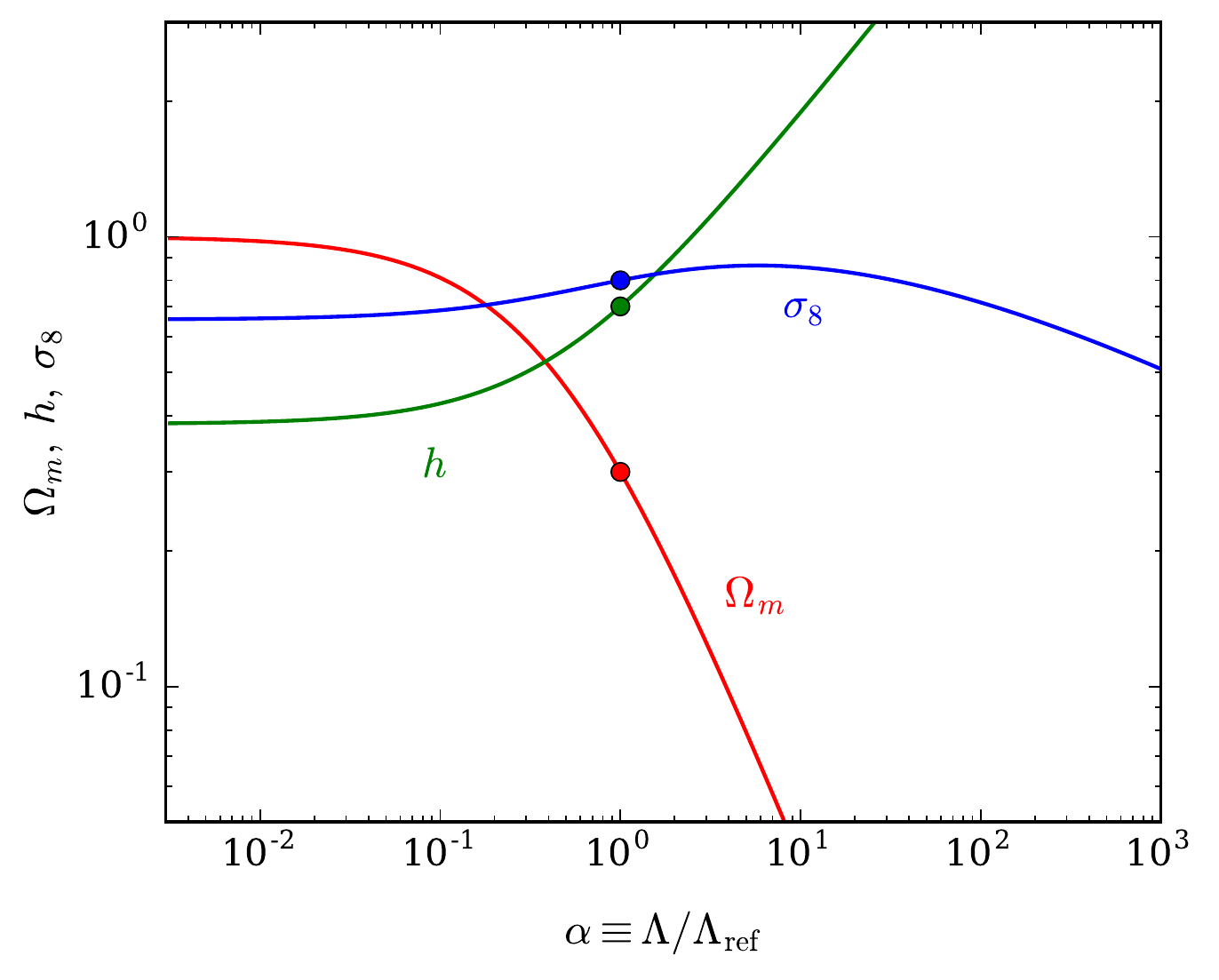}
	\caption{This plot shows how the principal $\Lambda$CDM parameters respond
to a scaling of the cosmological constant while maintaining the high-redshift universe otherwise unchanged, and exactly flat. The $z=0$ `present' is always defined as CMB temperature $2.725\,$K. Solid points show the default reference cosmology $(\Omega_m, h, \sigma_8)=(0.3,0.7,0.8)$, and how this scales as $\Lambda$ is altered from the reference value $\Lambda_{\rm ref}$ (which corresponds to $\Omega_\Lambda=0.7$).}
\label{fig:scale_lambda}
\end{figure} 

\begin{table*}
	\caption[List of simulations in Chapter \ref{chap:counterfactual}.]{List of simulations in this paper with their corresponding reference name from this point onwards. We have included $\Omega_m$, $\Omega_\Lambda$, $\Omega_b$, $\sigma_8$, $h$, cosmological box size of the simulations and starting $z$ of the simulations. All models assume a primordial scalar spectral index $n_s=0.96$. The box size of all simulations has been set to the same value in cMpc, independent of $h$. Refer to Section \ref{sec:cosmo-setup} for more information about the naming convention and specifics of each simulation.}
	\label{tab:diff-uni-setup}
	\begin{tabular}{|p{.11\linewidth}|p{.08\linewidth}|p{.08\linewidth}|p{.08\linewidth}|p{.08\linewidth}|p{.08\linewidth}|p{.14\linewidth}|p{.1\linewidth}|}
		\hline
		\multicolumn{8}{|c|}{Simulation setup list} \\
		\hline
		\rule{0pt}{10pt}Reference name & $\Omega_m$ & $\Omega_\Lambda$ & $\Omega_b$ & $\sigma_8$ & $h$ & Box size [$h^{-1}\,{\rm cMpc}$] & Starting $z$\\
		\hline
		{\sl LCDM} & 0.285 & 0.715 & 0.0461 & 0.828 & 0.695 & 50 & 99\\
		\hline
		{\sl EDS} & 1.000 & 0.000 & 0.1618 & 0.6786 & 0.371 & 26.69 & 25.69\\
		\hline
		{\sl 10x} & 0.0383 & 0.9617 & 0.0062 & 0.8742 & 1.895 & 136.34 & 99.027\\
		\hline
		{\sl 50x} & 0.0079 & 0.9921 & 0.00128 & 0.7805 & 4.172 & 300.14 & 99.04\\
		\hline
		{\sl 100x} & 0.0040 & 0.9960 & 0.00065 & 0.7217 & 5.8885 & 423.63 & 98.748\\
		\hline
		{\sl s5x} & 0.285 & 0.715 & 0.0461 & 0.484 & 0.695 & 50 & 99\\
		\hline
		{\sl s10x} & 0.285 & 0.715 & 0.0461 & 0.384 & 0.695 & 50 & 99\\
		\hline
		{\sl s20x} & 0.285 & 0.715 & 0.0461 & 0.305 & 0.695 & 50 & 99\\
		\hline
		{\sl s100x} & 0.285 & 0.715 & 0.0461 & 0.178 & 0.695 & 50 & 99\\
		\hline
	\end{tabular}
\end{table*} 
\section{Simulating counterfactual universes}\label{sec:sim_setup}

\subsection{Initial conditions}\label{sec:ic-scaling}

The simplest anthropic ensemble is a subset within the multiverse in which only the value of $\Lambda$ varies between the universes. All other key parameters, in the form of dimensionless numbers such as the photon-to-baryon ratio, are assumed to remain unchanged between members of the ensemble. In effect, we can imagine making multiple copies of the universe at high redshifts when $\Lambda$ is so small that its effect is non-existent; we can then scale $\Lambda$ up or down in these universes, which will have negligible effect until $\Lambda$ dominates at some later time. The amount of scaling will modify the cosmological parameters at $z=0$ -- which we define for convenience as the time when the temperature of the CMB is $2.725\,{\rm K}$. 


The altered $z=0$ cosmological parameters are of interest as they are the practical means by which we will carry out simulations of the counterfactual universes. The dependence on $\alpha$, the scaling factor for $\Lambda$, can be deduced as follows. First note that all present-day physical densities of the different cosmological constituents are unchanged:
\begin{equation}\label{eq:simple_omega_mh}
\tilde{\Omega}_j \tilde{h}^2= \Omega_j h^2,
\end{equation}
where a tilde denotes the modified value of the parameter, and the suffix $j$ can refer to radiation, CDM, or baryons (we assume that the neutrino mass is negligibly small). This arises because $\rho_j\propto\Omega_j h^2$, because defining $z=0$ at
fixed temperature fixes $\rho_r$, and because we fix  the baryon fraction and the number of photons per baryon.
In order obtain $\tilde h$ and $\tilde\Omega_j$ separately, we make appropriate use of the Friedmann equation, using the requirement that the expansion rate is not changed at early times where $\Lambda$ is negligible.
Suppose we scale $\Lambda$ by $\alpha$ at some early time when the scale factor is $a_i$, while maintaining spatial flatness. The subsequent evolution of the Hubble parameter is 
\begin{equation} \label{eq:hi_evo}
{\tilde H}^2=H_i^2[(1-\alpha\epsilon)(a_i/a)^3+\alpha\epsilon)],
\end{equation} where $\epsilon=\rho_v/\rho_{\rm tot}$ and $\rho_v$ is the vacuum density at $a_i$.  At early times, $\epsilon$ is small enough that we can set $(1-\alpha\epsilon)$ to unity. Comparing with the reference $\Lambda$CDM model, $H^2 = H_0^2(\Omega_ma^{-3}+ 1-\Omega_m)$, and setting ${\tilde H}=H$ at early times, this yields $H_i^2a_i^3=\Omega_mH_0^2$. Finally, the fiducial vacuum fraction at $a_i$ is 
\begin{equation}
\epsilon={(1-\Omega_m)\over(1-\Omega_m)+\Omega_m a_i^{-3}} \rightarrow {(1-\Omega_m)\over\Omega_m a_i^{-3}}\;{\rm for} \; a_i\ll1.
\end{equation}
Thus $\epsilon=(1-\Omega_m)H_0^2/H_i^2$ and 
we can express $\tilde{H}(a)$ in terms of our reference parameters,
\begin{equation}
\tilde{H}=H_0[\Omega_m a^{-3}+ \alpha(1-\Omega_m)]^{1/2},
\end{equation} 
which gives the $z=0$ Hubble parameter as
\begin{equation}
\tilde{h}=h[\Omega_m + \alpha(1-\Omega_m)]^{1/2}.
\end{equation}
This relation can be obtained directly from the Friedmann equation by noting that we do not change the matter density at $z=0$, but only scale the vacuum density. Since $\rho\propto\Omega H_0^2$, the result follows.
Because the total density remains critical, $\tilde{\Omega}_m$ is the matter density divided by the sum of matter and vacuum densities at $a=1$: $\Omega_m/[\Omega_m+\alpha(1-\Omega_m)]$, or
\begin{equation}
(1-1/\tilde{\Omega}_m)=\alpha(1-1/\Omega_m).
\end{equation}
This recasting of the cosmological parameters is reminiscent of the `separate universe' technique common in numerical simulations (e.g. \citejap{2015MNRAS.448L..11W}), but with the critical difference that in the latter case the curvature is allowed to vary. Here the assumption is that whatever generates the multiverse ensemble, such as inflation, has a mechanism that guarantees flatness in all cases.

Finally, we also need the normalization $\tilde{\sigma}_8$ in this modified cosmology. If the comoving filter length in the reference $\Lambda$CDM model is $R=8\, h^{-1}\,$cMpc, then $\sigma(R)$ needs to have the same value at high redshift in all models. However, $\sigma(R)$ at $z=0$ when the temperature of the CMB is 2.725 K is now altered because the modification of the value of $\Omega_m$ will change the linear growth factor \citep{2007APh....28..481L}. Furthermore, $R$ is no longer $8\mpcoh$ in terms of $\tilde{h}$, but we wish by convention to specify the normalization in terms of $\tilde\sigma_8$, which is the $z=0$ value of $\sigma$ with a filtering radius of $8\,\tilde{h}^{-1}{\rm cMpc}$. We therefore
need to scale the $z=0$ $\sigma(R)$ by a factor that depends on the shape of the power spectrum in order to specify a normalization that keeps the early universe physically identical in all cases.  The results of this exercise, together with the dependence of $\tilde{\Omega}_m$ and $\tilde{h}$ on $\alpha$, are shown in Figure~\ref{fig:scale_lambda}. 

For any value of $\alpha$, we can thus obtain an altered set of $z=0$ cosmological parameters, which can then be used to generate cosmological simulations with an unaltered $\Lambda$CDM simulation code. The only complication is the sign of $\alpha$: the scaling formulae are general, and will yield correct results if the high-$z$ cosmological constant is replaced by a negative value, but complications can arise in the subsequent evolution. With $\Lambda<0$, the cosmological expansion will eventually reach a maximum and then undergo recollapse. A standard $N$-body code is likely to fail at this point, since the relation between time and scale factor is normally assumed to be monotonic. Moreover, if $|\Lambda|$ is sufficiently large, the recollapse will occur very early: the minimum temperature will be above $2.725\,$K, so that $z=0$ will not be reached. We can see this from the above expressions, in which $\tilde h$ is undefined for $\alpha< -1/\Omega_m+1$. These complications are not too difficult to deal with, but we do not address them in this paper, which is concerned purely with the astrophysical impact of a large positive $\Lambda$ in freezing out structure growth and subsequent star formation.

As an alternative to these scaled-$\Lambda$ models, we also conduct a set of simulations with a simpler parameter variation in which we alter only the  normalization $\sigma_8$, while retaining unchanged the values of the other $\Lambda {\rm CDM}$ parameters. The rationale for this is that a principal effect of raising the value of $\Lambda$ is to cause earlier freezeout and hence yield a lower value of $\sigma_8$. This reduced normalization is the principal reason for the operation of Weinberg's anthropic argument, since it causes a corresponding reduction in the abundance of galaxy-scale dark-matter haloes. It is therefore of interest to look at the effect of a direct change in normalization, as opposed to a change induced as a consequence of changing $\Lambda$. For a given value of $\sigma_8$ in the far future, to what degree will the results depend on the evolutionary track by which it was attained? This question motivates us to scale $\sigma_8$ in order to obtain an equivalent combination of the cosmological parameters to one where $\Lambda$ is changed. We know that $d\ln \delta/ d\ln a \approx \Omega_m^{0.55}$ \citep{2007APh....28..481L} and growth freezes when $\Omega_m = \Omega_\Lambda = 0.5$. Suppose we take a universe where we increase $\rho_\Lambda$ by a factor $\alpha$: the scale factor where $\rho_\Lambda$ = $\rho_m$ is then reduced by a factor $\alpha^{1/3}$. Since fluctuations grow linearly with $a(t)$ until this point, we therefore reduce the value of $\sigma_8$ by the same factor of $\alpha^{1/3}$ relative to the reference $\Lambda \rm CDM$ universe.

\subsection{Simulation setup} \label{sec:cosmo-setup}

The scaling relations from Section \ref{sec:ic-scaling} allow us to produce sets of cosmological parameters for models where we change either the value of $\Lambda$ or $\sigma_8$ from a reference $\Lambda \rm CDM$ universe. The chosen values are summarised in Table \ref{tab:diff-uni-setup}. The reference $\Lambda {\rm CDM}$ cosmological parameters are adopted from WMAP-9 \citep{2013ApJS..208...20B}:
\begin{equation}
\eqalign{
&(\Omega_m, \,\Omega_\Lambda, \,\Omega_b, \,h, \,\sigma_8,n_s)= \cr
&(0.285, \,0.715, \,0.0461, \,0.695, \,0.828, \,0.96),
}
\label{eq:cosmo_para}
\end{equation}
where the parameters have their usual definitions. We generate the initial conditions for these universes with MUlti-Scale Initial Conditions ({\tt MUSIC}) for cosmological simulations \citep{2011MNRAS.415.2101H} and evolve them with {\tt Enzo} \citep{2014ApJS..211...19B}, using the hydrodynamic solver {\tt ZEUS} \citep{1992ApJS...80..753S}, and an N-body adaptive particle-mesh gravity solver \citep{1985ApJS...57..241E}. We also couple the simulation with the {\tt Grackle} gas-phase chemistry package \citep{2017MNRAS.466.2217S}. We used this simulation apparatus in paper 1: \citet{2020MNRAS.497.5203O}, a study of cosmic star formation focusing on galaxies similar to the Milky Way, where we found that reliable results to $z=0$ for a fiducial {\sl LCDM} simulation could be achieved using a base mesh of $128^3$ and four levels of AMR refinement, yielding a spatial resolution of 35\,ckpc and a mass resolution of $6.8\times 10^9\,M_\odot$. In paper 2: \citet{2021MNRAS.507.5432O}, we extended this calculation into the future; this required a number of modifications to the standard code, which we include here.

As listed in Table \ref{tab:diff-uni-setup}, we consider scaling factors for $\Lambda$ of $\alpha = 0, 10, 50, 100$, yielding simulations labelled as {\sl EDS}, {\sl 10x} and {\sl 100x} respectively. The value of $\alpha=0$ implies a flat universe fully dominated by matter, the Einstein--de Sitter (EdS) model \citep{1932PNAS...18..213E}. We also reduce $\sigma_8$ by a factor of $\beta^{1/3}$, where $\beta=5, 10, 20, 100$. These simulations are named {\sl s$\beta$x} with a `s' prefix to distinguish them from the simulations with scaled $\Lambda$. 

In addition to the cosmological parameters, Table \ref{tab:diff-uni-setup} includes two additional columns, giving the comoving box size and starting redshift used to generate the initial conditions for the counterfactual universes. For {\sl LCDM}, the mass and maximum spatial resolutions are identical to simulation {\sl NL} in \citet{2021MNRAS.507.5432O}. The comoving box size is specified in units of $\mpcoh$, but we change this value according to $h$ in different cosmologies so that it has a fixed value in cMpc, in order to maintain a consistent spatial resolution and an identical particle mass across all simulations.  This issue does not arise for simulations with scaled $\sigma_8$, since $h$ is the same for all these models; but for {\sl EDS}, {\sl 10x} and {\sl 100x}, $h$ differs significantly between the universes. 

The starting redshift of the simulations is also adjusted such that the universes are at the same starting age. We can convert the starting redshift of 99 in the fiducial model to a corresponding value for the counterfactual universes by using the relation
\begin{equation}
\label{eq:z_vs_t}
H_0t = \frac{2}{3\sqrt{1-\Omega_{m}}}\sinh^{-1}\sqrt{\frac{1-\Omega_{m}}{\Omega_{m}(1+z)^3}}\;, 
\end{equation} where $\mathrm{H_0}$ and $\Omega_{m}$ is the Hubble parameter and matter density parameter at $z=0$ \citep{1993ppc..book.....P}. Again, we do not need to change the starting redshifts for simulations with scaled $\sigma_8$. For {\sl EDS}, {\sl 10x} and {\sl 100x}, the starting redshifts are 25.69, 99.03 and 98.75 respectively.

\begin{figure*}
	\centering
	\subfloat[EDS\label{fig:pp-0x}]{%
		\includegraphics[width=.5\linewidth]{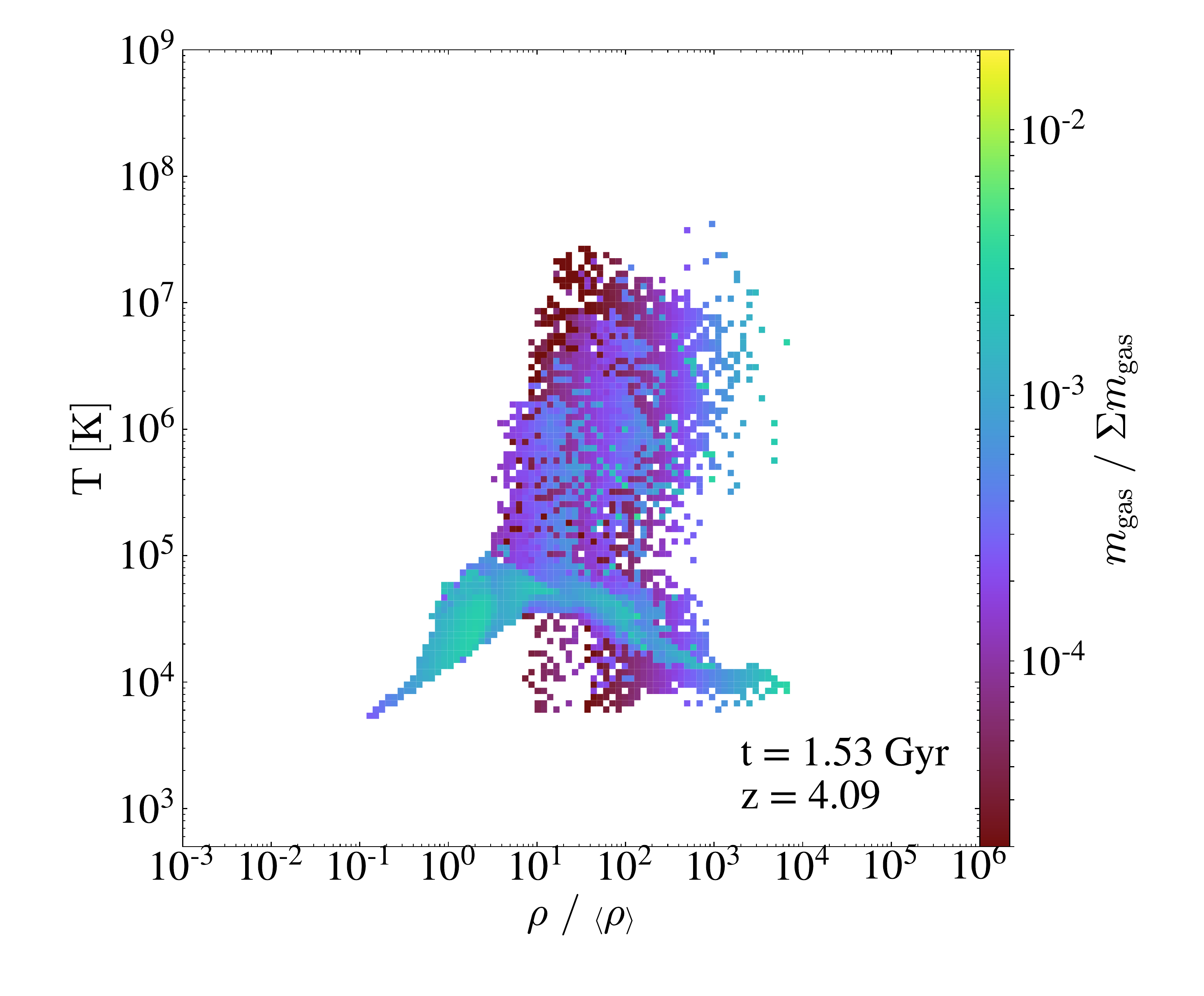}
	}
	\subfloat[LCDM\label{fig:pp-1x}]{%
		\includegraphics[width=.5\linewidth]{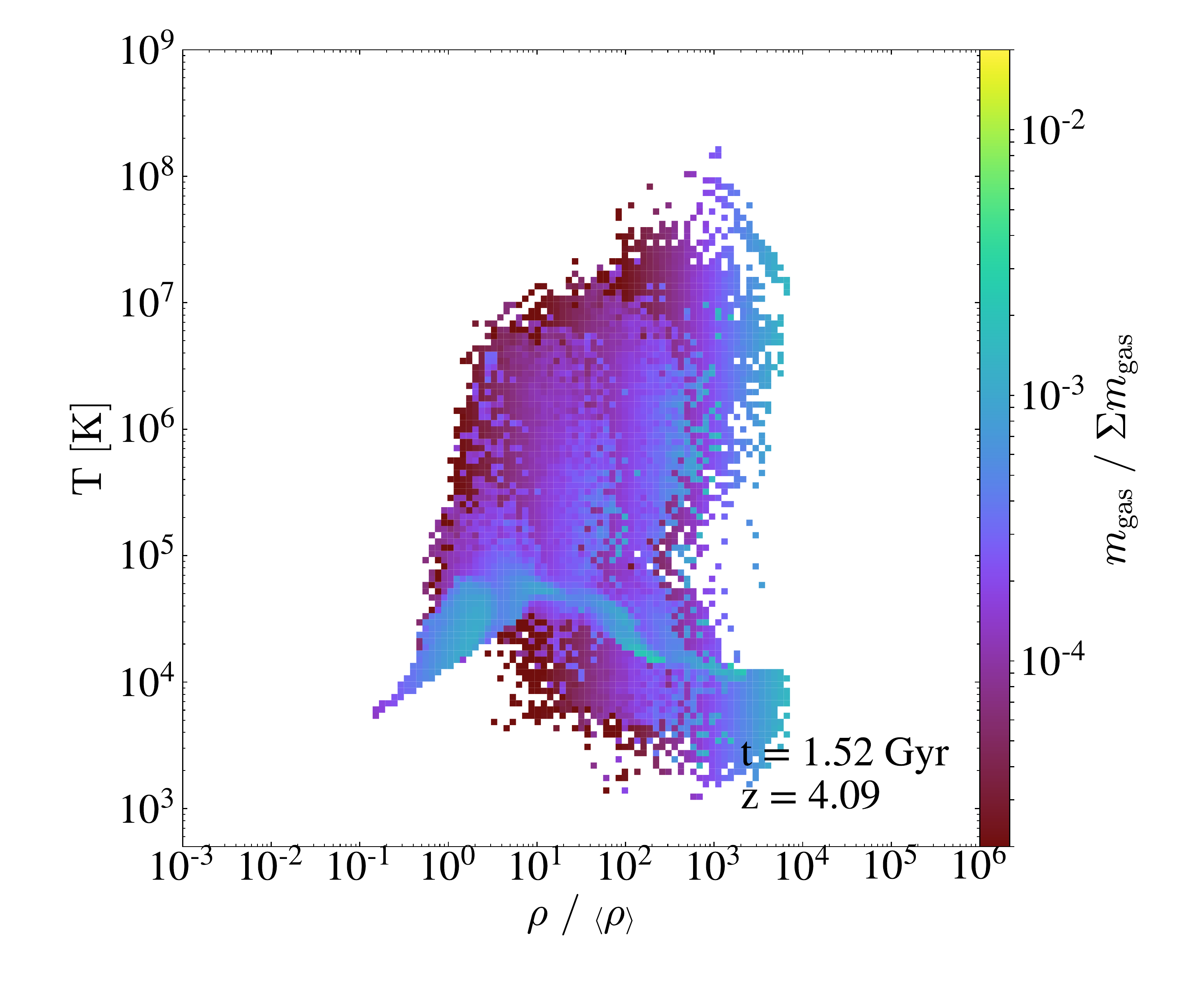}
	}
	\hfill
	\subfloat[10x\label{fig:pp-10x}]{%
		\includegraphics[width=.5\linewidth]{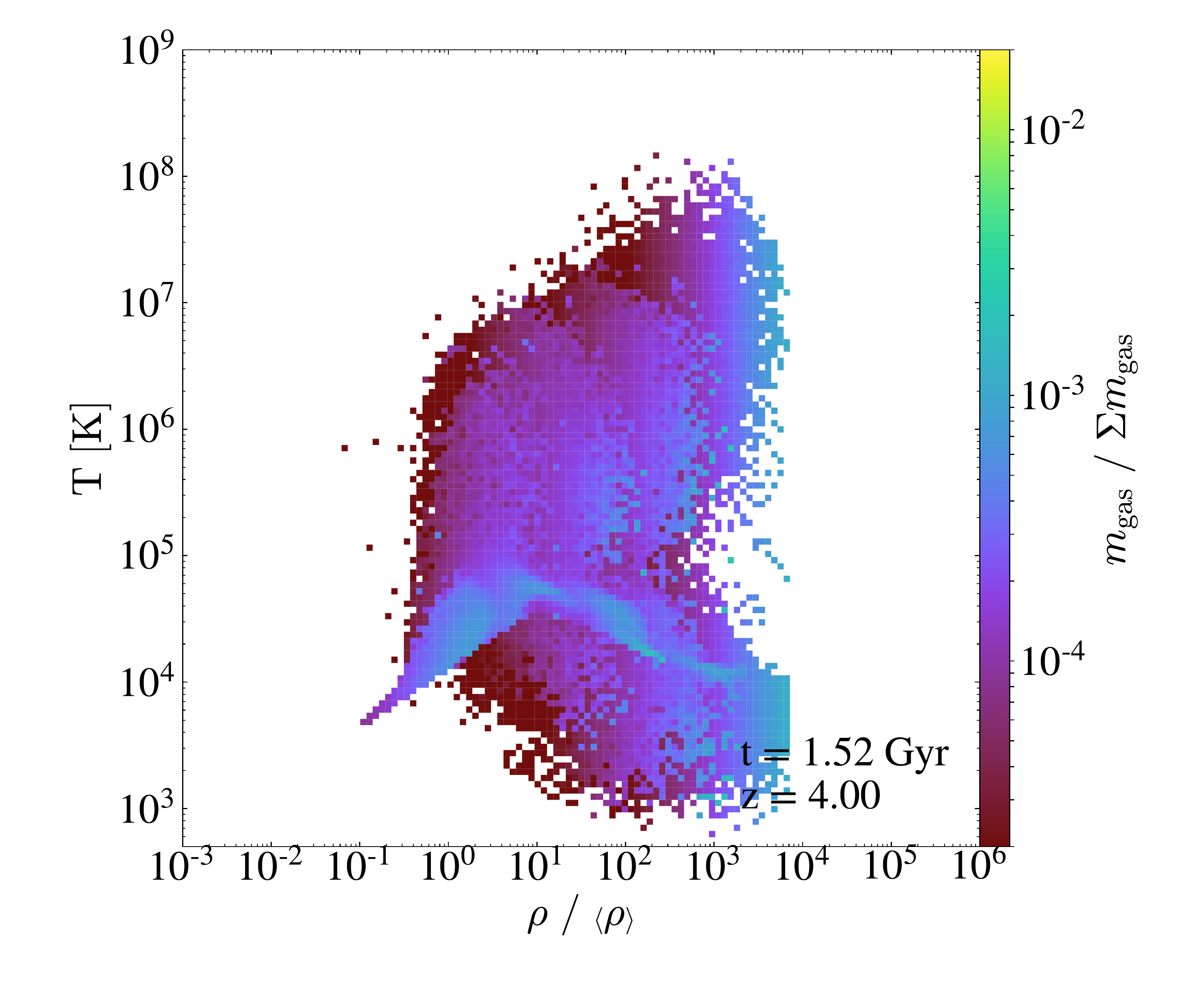}
	}
	\subfloat[100x\label{fig:pp-100x}]{%
		\includegraphics[width=.5\linewidth]{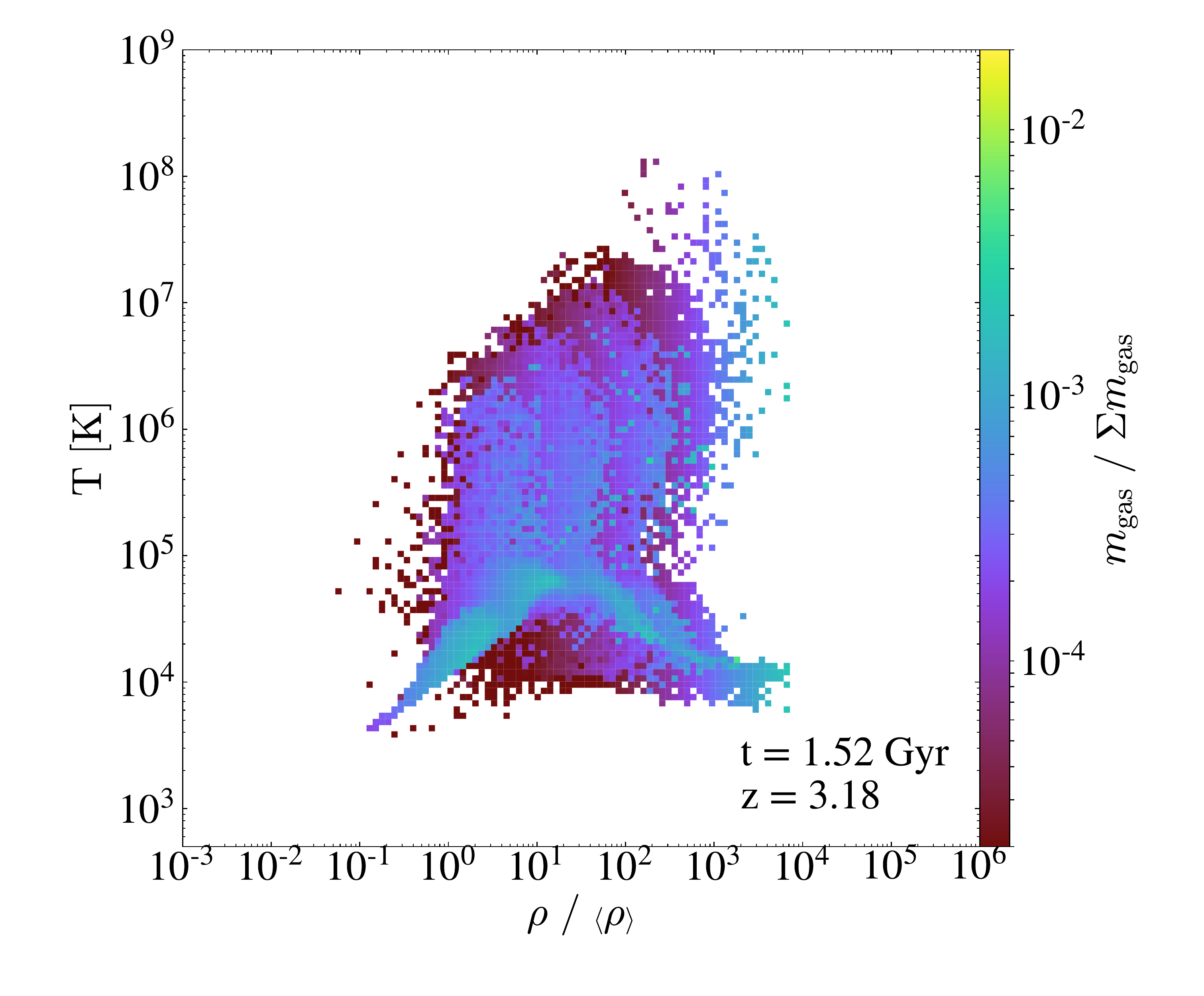}
	}
	\caption{Gas temperature against gas overdensity at $t\approx1.5\,{\rm Gyr}$ for the simulations of different $\alpha$ indicated in the captions. The colour bar indicates the gas mass in a given pixel of temperature and overdensity, expressed as a fraction of the total gas mass within the simulation volume. The general features of the plots are similar to each other at this time, validating our assignment of modified cosmological parameters, which are chosen so that all alternative models should have the same evolution in time, until $\Lambda$ becomes dominant. 
	}
	\label{fig:alpha-pp}
\end{figure*} 

\subsection{Baryonic physics}

Our simulations adopt the \citet{1992ApJ...399L.113C} and \citet{2006ApJ...650..560C} models for star formation and feedback respectively, with modifications made by \citet{2011ApJ...731....6S}.  An extensive parameter study and calibration of these algorithms within {\tt Enzo} has been carried out by \citet{2020MNRAS.497.5203O}. We provide here a short summary of the parameters involved in the modelling.

Out of the two setups explored in \citet{2020MNRAS.497.5203O}, we focus on the application of `Setup 2', which employs a timestep-independent conversion of gas into stars. According to the calibrated results, we set $\epsilon=3.0\times10^{-5}$, $r\_s=1\_1$ and $f_*=0.9$, where $\epsilon$ is a factor that relates the amount of feedback energy to the rest mass energy of the star forming gas (see Equation 6 in \citejap{2020MNRAS.497.5203O}); $r\_s$ specifies the volume into which the feedback energy is deposited (6 adjacent cells to the star particle for $r\_s=1\_1$); and $f_*$ indicates the conversion efficiency of gas mass in a cell into stellar mass (see sections 2.1 and 2.2 in \citejap{2020MNRAS.497.5203O} for details). This set of parameters successfully reproduced the baryonic makeup of a Milky-way sized halo at $z=0$, as quantified in the analysis by \citet{2010ApJ...708L..14M}.

Extending the study of the impact of feedback physics on the evolution of baryonic properties, \citet{2021MNRAS.507.5432O} evolved the same setup beyond $z=0$. Such a setup in a standard $\Lambda$CDM universe forms a similar total stellar mass by $z=-0.995$ (approximately $96\,{\rm Gyr}$) to that predicted by extrapolating the analytical SFRD fit from \citet{2014ARA&A..52..415M}. This agreement is also achieved regardless of the resolution of the simulation.

The starting assumption of the current work is that the above modelling correctly represents the operation of the full baryonic physics that applies on scales below those that can be resolved in the simulation, and that
this same small-scale modelling can be used in universes beyond the $\Lambda$CDM case that was used for calibration.  To the extent that the parameters of the modelling refer to small-scale local quantities, this should be a defensible assumption. Furthermore, we consider only simple counterfactual cosmologies in which the densities of baryons and dark matter are always in the same ratio, so that the internal dynamics of a halo should be followed self-consistently by {\tt Enzo} in a manner that is independent of the large-scale cosmological context. In any case, it is certainly of interest to see how the modelling functions in universes very different from our own. In the absence of observations, the robustness of such predictions can only be tested by comparison with alternative calculations that use different subgrid approaches. But here we can do no more than explore the predictions of the existing {\tt Enzo} model, taking care to ensure that the results are at least converged within our resolution limits.

\section{The long-term evolution of the halo mass function} \label{sec:evol-hmf-igm}

In this section, we will compare various properties of both gas and dark matter in the simulations, taking a similar line to the comparison that was carried out in \citet{2021MNRAS.507.5432O}. But now focusing on the way in which cosmology (specifically $\Lambda$) affects the picture of structure formation and evolution. 

We first validate the different initial conditions in our simulations. The counterfactual $z=0$ cosmology parameters have been adjusted so that the evolution of the universe in its early matter-dominated phase should be identical for all models. We illustrate this test of our simulations in Figure \ref{fig:alpha-pp}: this presents a phase plot, which shows how the gas in different cosmologies is distributed in the temperature-baryon overdensity phase space. Panels (a), (b), (c) and (d) correspond to {\sl EDS}, {\sl LCDM}, {\sl 10x} and {\sl 100x} respectively.
We see that the general distribution of gas in the temperature-overdensity plane is indeed very similar for all these models. The agreement is not perfect, because we have chosen a relatively late output time in order to capture significant nonlinear evolution in the IGM: $t=1.09\,{\rm Gyr}$ in all cases (although the nearest output for {\sl EDS} differs slightly from this figure). This time translates to $z=5.36$ in {\sl LCDM}, at which point $\Lambda$ is not so far from becoming significant, especially for the models where its value is scaled up substantially.

\subsection{Halo mass function} \label{sec:new_cosmo-hmf}

\begin{figure*}
	\centering
	\subfloat[$t \approx 0.5\ t_{\rm H}$]{%
		\includegraphics[width=.5\linewidth]{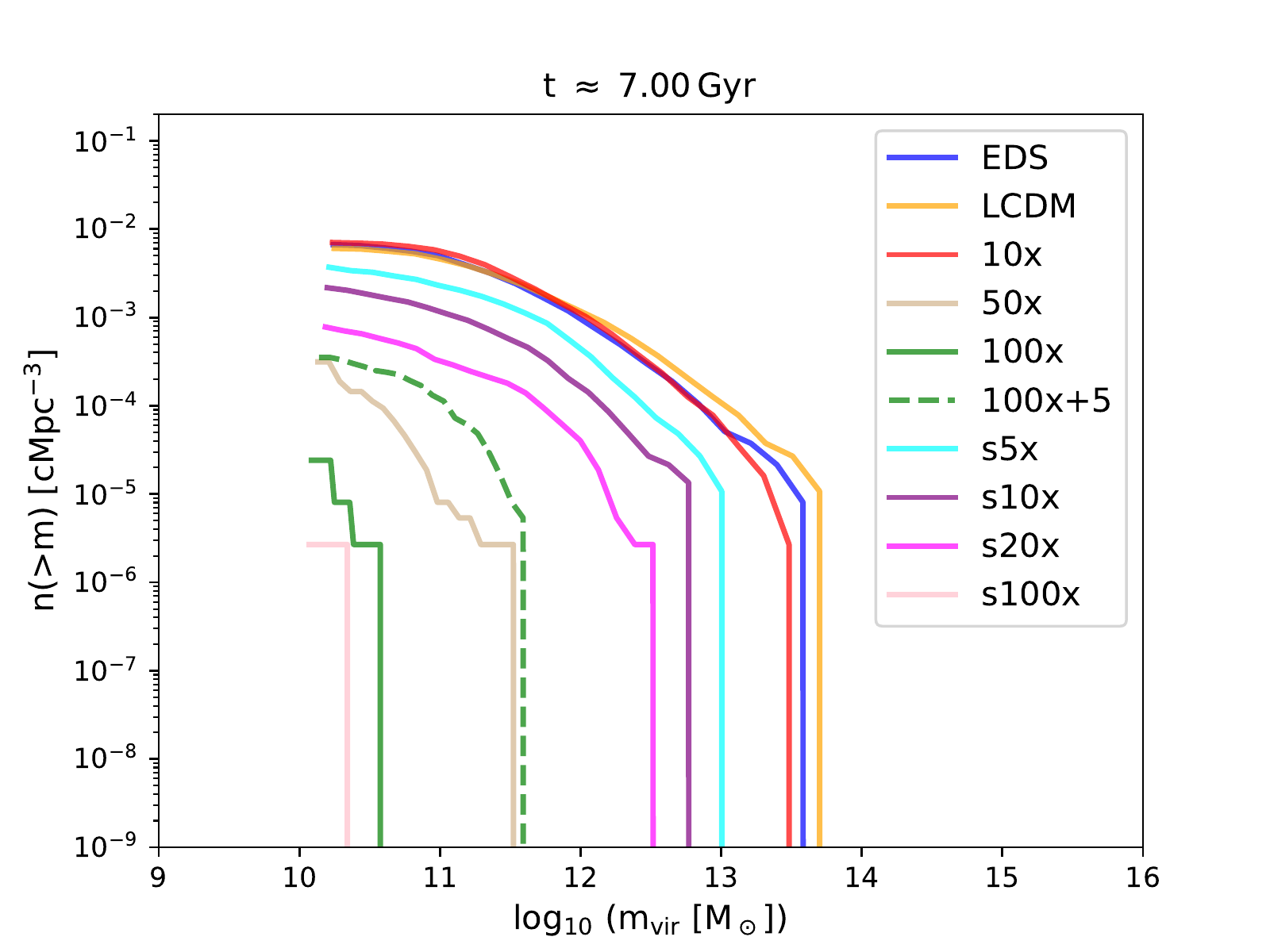}
	}
	\subfloat[$t \approx t_{\rm H}$]{%
		\includegraphics[width=.5\linewidth]{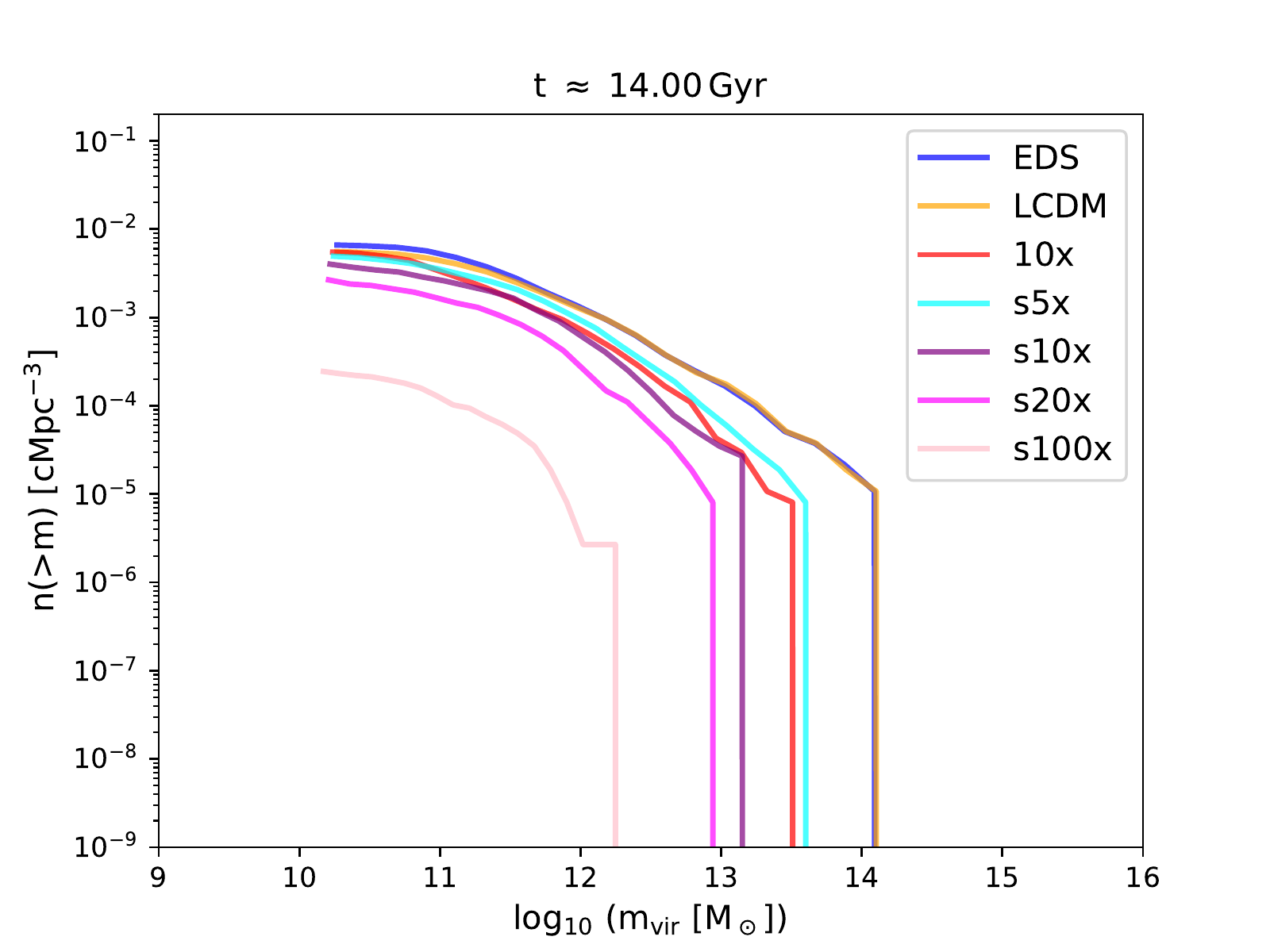}
	}
	\hfill
	\subfloat[$t \approx 2\ t_{\rm H}$]{%
		\includegraphics[width=.5\linewidth]{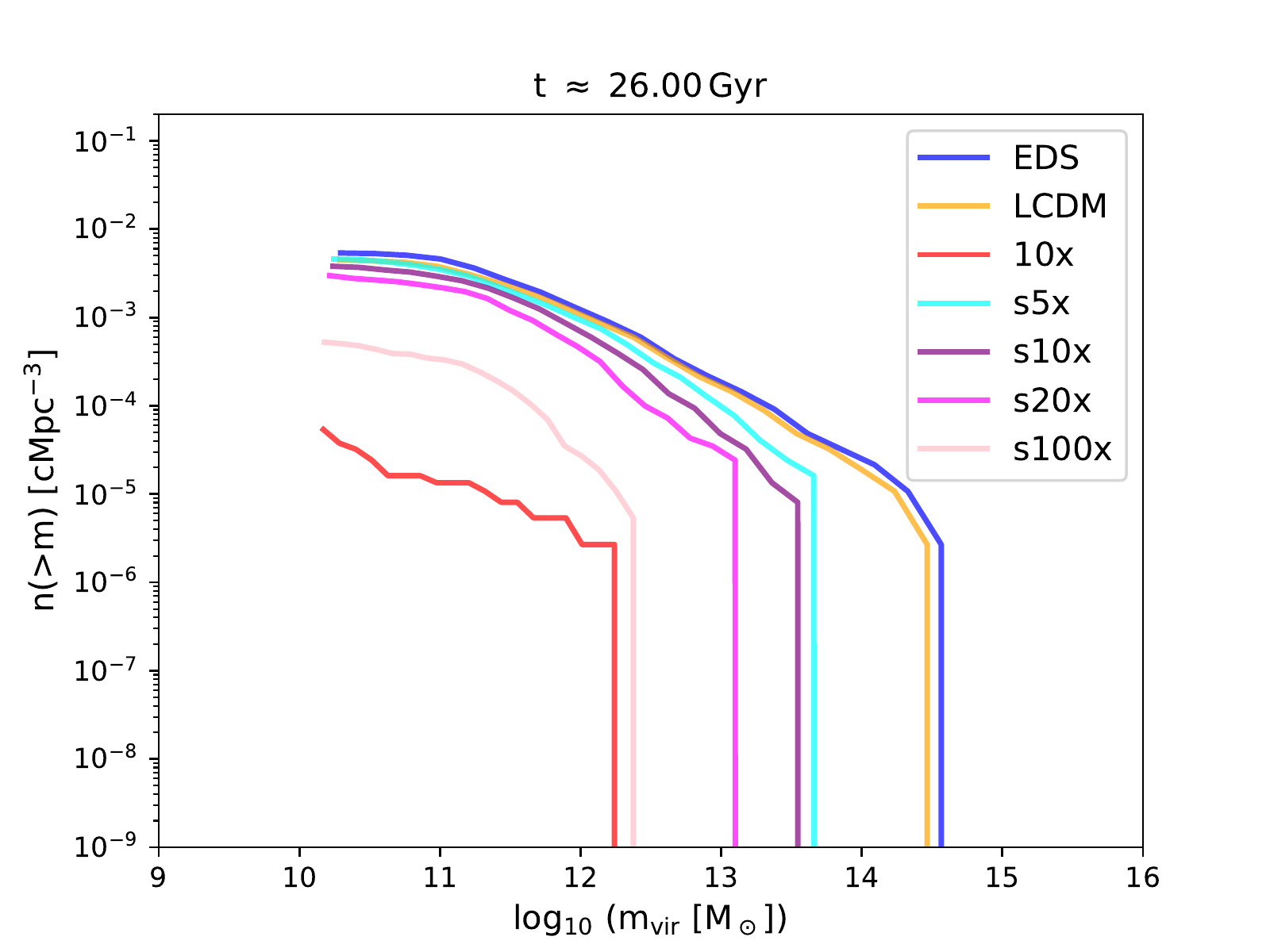}
	}
	\subfloat[$t \approx 4\ t_{\rm H}$]{%
		\includegraphics[width=.5\linewidth]{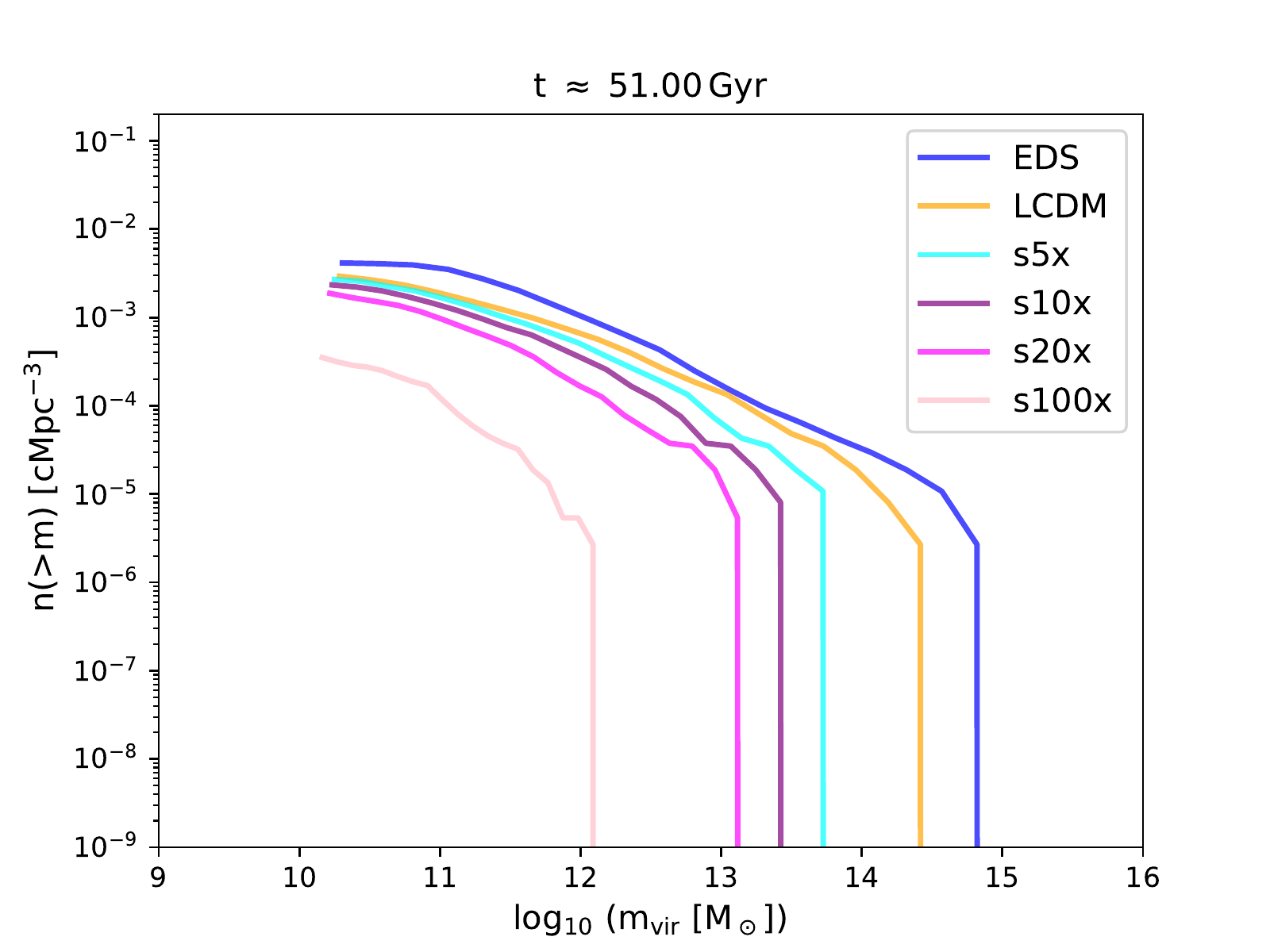}
	}
	\caption{Time evolution of the HMF from the simulations of our different cosmologies. The blue, red, tan, green, cyan, purple, magenta, pink and orange lines refer to {\sl EDS}, {\sl 10x}, {\sl 50x}, {\sl 100x}, {\sl s5x}, {\sl s10x}, {\sl s20x}, {\sl s100x} and {\sl LCDM} respectively. The general evolution is that the HMF will advance across the plot from the left with little change in shape or normalization of the curve, before ceasing to evolve at a time corresponding to $\Lambda$ dominance. This is as expected from simple analytic arguments; but subsequent evolution departs from this behaviour for highly scaled models, with the mass function reversing its evolution and  `exiting' the plot to the left. The onset of this unphysical behaviour reflects the point at which the resolution of the simulation ceases to be adequate to follow the population of haloes, which shrink without limit in comoving coordinates as the universe continues to expand. There is thus a tendency for haloes to become unbound and lost from the halo catalogue (defined with a minimum size of 20 particles). \JAP{In order to understand this effect, we were able to repeat our simulation with largest $\Lambda$ with increased resolution, and this is shown as the alternative green dashed line in the $t=7\,$Gyr panel.} Further discussion of this resolution issue is given in Section \ref{sec:new_cosmo-hmf}.
	}
	\label{fig:new_cosmo-hmf}
\end{figure*}

In this section, we test the sensitivity of the Halo Mass Function (HMF) to cosmology, using halo catalogues obtained via the {\tt ROCKSTAR} halo finder \citep{2013ApJ...762..109B}, in which each halo is resolved by a minimum of 20 particles. In Figure \ref{fig:new_cosmo-hmf} we display the evolution of the HMF in our different universes. We choose several snapshots at $t \approx 7,\ 14,\ 26,\ 51\,{\rm Gyr}$, which correspond roughly to $0.5\ t_{\rm H},\ t_{\rm H},\ 2\ t_{\rm H}$ and $4\ t_{\rm H}$ with $t_{\rm H} \approx 13.7\,{\rm Gyr}$ in panels (a) to (d) respectively. Each coloured line refers to the same cosmology consistently across the panels as indicated in the legend and caption. 

\begin{figure*}
	\centering
	\includegraphics[width=0.8\linewidth]{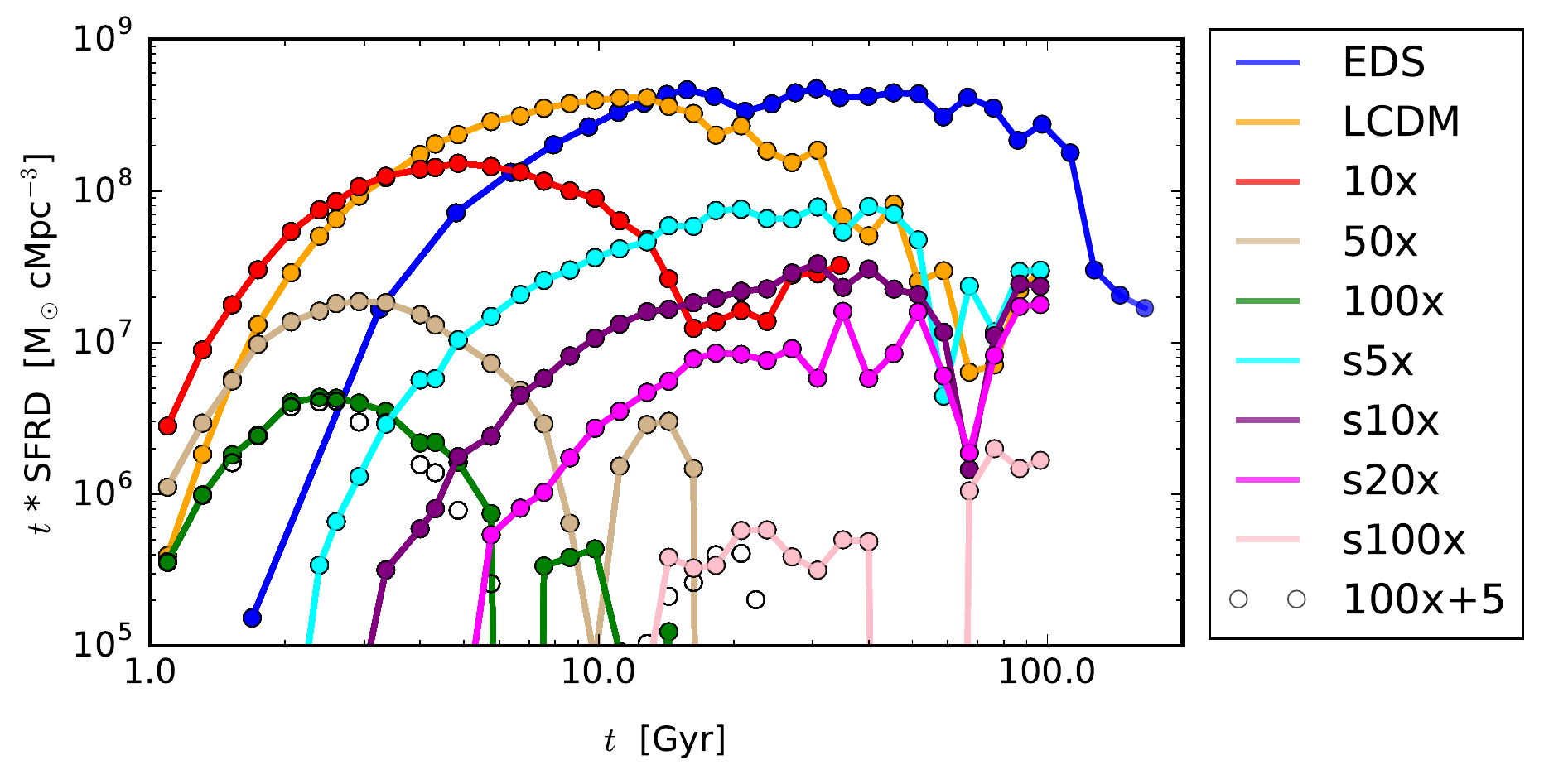}
	\caption{Evolution of the SFRD for counterfactual universes across cosmic time. We use the same colours as in Figure \ref{fig:new_cosmo-hmf} to represent the different cosmologies. \JAP{For {\sl $100x$} we show results at two maximum AMR levels, with open circles indicating the higher resolution.} The evolution of the cosmic SFRD differs significantly between the universes. Note that $z = 0$ is defined as the time when the temperature of the CMB is 2.725\,K, and  comoving lengths are defined relative to this era. We discuss specific differences of the SFRD arising from the cosmology in Section \ref{sec:evol-sfr}.}
	\label{fig:new_cosmo-sfr}
\end{figure*} 
Since structure formation in the CDM paradigm happens in a bottom-up manner, low mass haloes form first from the primordial density fluctuations before merging to create more massive haloes. Thus the HMF `enters' the plot in Figure \ref{fig:new_cosmo-hmf} from the low mass end. The characteristic cutoff in the HMF then moves towards higher mass  as time increases, and is expected to saturate at an asymptotic form once $\Lambda$ dominates. However, the continuing expansion of the universe induces a deterioration of the {\it proper\/} force resolution in our simulations. Once this resolution scale exceeds the virial radius of a halo, the particles in the halo experience too small a gravitational force to remain bound, causing the haloes to dissolve and fade into the background. This eventual loss of haloes is apparent on all mass scales: see e.g. the red line in Figure \ref{fig:new_cosmo-hmf}, where the mass function for {\sl 10x} is seen `exiting' the plot to the left at $t=26$\,Gyr, being strongly suppressed with respect to its value at $t=14$\,Gyr.
It may seem puzzling that the mass function is not eroded first at the lowest masses: these have the smallest virial radii and hence should arguably be the first to experience the effects of insufficient resolution. However, all haloes have a central spike in density where the velocity dispersion is highest, and declining proper resolution will lead to particles being able to escape from this region, so that even the most massive haloes dissolve, in an inside-out manner.

The same general pattern of evolution is seen for different cosmologies, although the unphysical turnaround of the HMF occurs at different times, depending on $\alpha$. With a larger $\alpha$, $\Lambda$ dominates earlier in the universe, leading to an earlier onset of exponential expansion and freezeout of growth. As a consequence, the resolution-induced decline in the number density of haloes across the mass range sets in earlier when $\alpha$ is increased. This difference is obvious when we compare {\sl 10x} (red), {\sl 100x} (green) and {\sl LCDM} (orange) in Figure \ref{fig:new_cosmo-hmf}. In panel (a) of that figure, we see that {\sl EDS}, {\sl LCDM}, and {\sl 10x} are all very similar in their mass functions, whereas {\sl 100x} is displaced to low masses, reflecting the fact that this model is already heavily $\Lambda$-dominated even at $t=7$\,Gyr. In turn, {\sl 10x} is separated from {\sl LCDM} in panel (b), because this model has already frozen out by $t=14$\,Gyr; and by $t=26$\,Gyr this model experiences resolution-induced loss of haloes. However, even by the latest time shown (51\,Gyr), there is no sign of such problems in {\sl LCDM} or {\sl EDS}. For the latter model, the slower expansion means that the force resolution remains adequate at all simulated times, and the lack of freezout means that the most massive halo in the simulation continues to gain mass as structure formation proceeds. For {\sl LCDM}, we expect that there will also in due course be a loss of haloes as exponential expansion fully sets in, but at $t=51$\,Gyr the expansion since $t=14$\,Gyr is a factor of 9.9 or 2.4 for {\sl LCDM} or {\sl EDS} respectively, so the proper resolutions of these models are not yet vastly different.

We can contrast this behaviour with what is seen when we simply scale the normalization $\sigma_8$ within $\Lambda {\rm CDM}$, with otherwise fiducial parameter values. 
If we neglect issues of numerical resolution, the main factor is a timing issue of when freezeout occurs in each counterfactual universe. For universes with scaled $\Lambda$, e.g., {\sl 10x} and {\sl 100x}, freezeout of the HMF into an unevolving asymptotic form will occur at increasingly early times, the more $\Lambda$ is increased.  For universes with unchanged $\Lambda$, we can scale $\sigma_8$ appropriately to yield this same asymptotic HMF.  We would then expect the differences between {\sl $\alpha$x}, {\sl s$\beta$x}, and {\sl LCDM} to be small once both models have frozen out, until resolution effects come to dominate. This is broadly the impression given by Figure \ref{fig:new_cosmo-hmf}. For example, in panel (b) of that plot, {\sl s10x} has a similar HMF to that of {\sl 10x}, at a time when the latter model has completely frozen out and the former very nearly so. But as discussed earlier, a larger $\alpha$ accelerates the expansion of the universe leading to a deterioration of the proper force resolution. Consequently, by $t=51.00\,{\rm Gyr}$, panel (d) displays HMFs only for {\sl s$\beta$x} models in addition to {\sl LCDM} and {\sl EDS}, and the halo populations of models with increased $\Lambda$ have not been retained to this time.

\JAP{This inability to follow the halo population into the indefinite future is a potential concern for our principal aim, which is to look at the long-term efficiency of star formation in counterfactual universes. This issue is discussed in the following Section, where we find that in all cases there is an era of peak star formation, followed by a decline in activity that shows good evidence for a convergence in the total production of stars. However, it is important to be convinced that this convergence is not an artefact caused by the disappearance of haloes within the simulation. In order to probe this issue, we chose to repeat the most strongly affected simulation, which is the universe with the largest $\Lambda$ ({\sl $100x$}). Here, we were able to achieve five additional AMR levels beyond the default of four, which was motivated by our previous results on the evolution to $z=0$ in {\sl LCDM}. Results from this run are contrasted with the base resolution in Figure~\ref{fig:new_cosmo-hmf}, where the loss of haloes apparent in {\sl $100x$} at 7\,Gyr is now removed. We consider below the impact of this variation in resolution on the star-formation history.}

\section{The long-term history of star formation} 
\label{sec:evol-sfr}

\subsection{counterfactual SFRDs and fits}

We now investigate how the long-term SFRD in these counterfactual models is influenced by cosmology. The maximum time into the future that can be investigated is limited by a number of factors: computational cost; onset of resolution systematics; and on occasion simple crashes of the code in extreme parameter regimes, as discussed in papers 1 \& 2. These limits are different for each simulation, as discussed below. But we believe that the star formation in each model shows good evidence of convergence by the point of its maximum reliable time.

Figure \ref{fig:new_cosmo-sfr} shows how the SFRDs evolve in our different simulated counterfactual universes, and a wide diversity of behaviour is apparent. Each cosmology displays a different peak in the SFR, both in terms of the time at which it occurs and the value at the peak. For a universe with higher $\alpha$ scaling applied to $\Lambda$, the peak in the SFRD happens earlier and reaches a lower maximum value. It is interesting to consider the main qualitative drivers of this behaviour. With a higher value of $\alpha$, $\Lambda$ dominates earlier in the universe, so that the HMF freezes out at a lower characteristic halo mass (See Section \ref{sec:new_cosmo-hmf}). In this exponentially expanding de Sitter phase, we expect that accretion of gas onto haloes will cease, leading ultimately to a reduction in star formation.

But there is also a peak in the SFRD for {\sl EDS}, in which there is no freezeout, consistent with the findings of \citet{2018MNRAS.477.3744S}. This behaviour is a reflection of gravitational heating of the IGM by continuous infall of matter and orbiting sub-structures, which convert gravitational potential energy into heat \citep{2008ApJ...680...54K,2008MNRAS.383..119D}. Thus baryonic physics must also be playing an important role in the SFRD peak.
At early times, there is general agreement that star formation is driven by the infall of new cold material from the cosmic web, rather than coming from the pre-existing baryons in haloes, which are driven into a hot gaseous halo
\citep{2009Natur.457..451D,2009ApJ...703..785D, 2015ARA&A..53...51S}. The question is then what happens to this material in the very long term, when it will eventually be able to cool and potentially generate further star formation. This is in effect the issue that is addressed directly by our simulations.

For counterfactual universes with reduced $\sigma_8$, structure formation and its associated star formation is  suppressed with respect to {\sl LCDM} in all cases.
However, the detailed effect is very different from when we scale $\Lambda$. As demonstrated by Figure \ref{fig:new_cosmo-sfr}, the peak in the SFRD for e.g. {\sl s10x} occurs much later than for the matched {\sl 10x} simulation, and the peak is much lower, so that the total amount of star formation in {\sl 10x} is higher than in {\sl s10x}, even though the activity in the latter case occurs over a larger range of time.

\begin{figure*}
	\centering
	\subfloat[Visual representation of fits to the simulated SFRD($t$) data. \label{fig:conv-sfrd-fit}]{%
		\includegraphics[width=.8\linewidth]{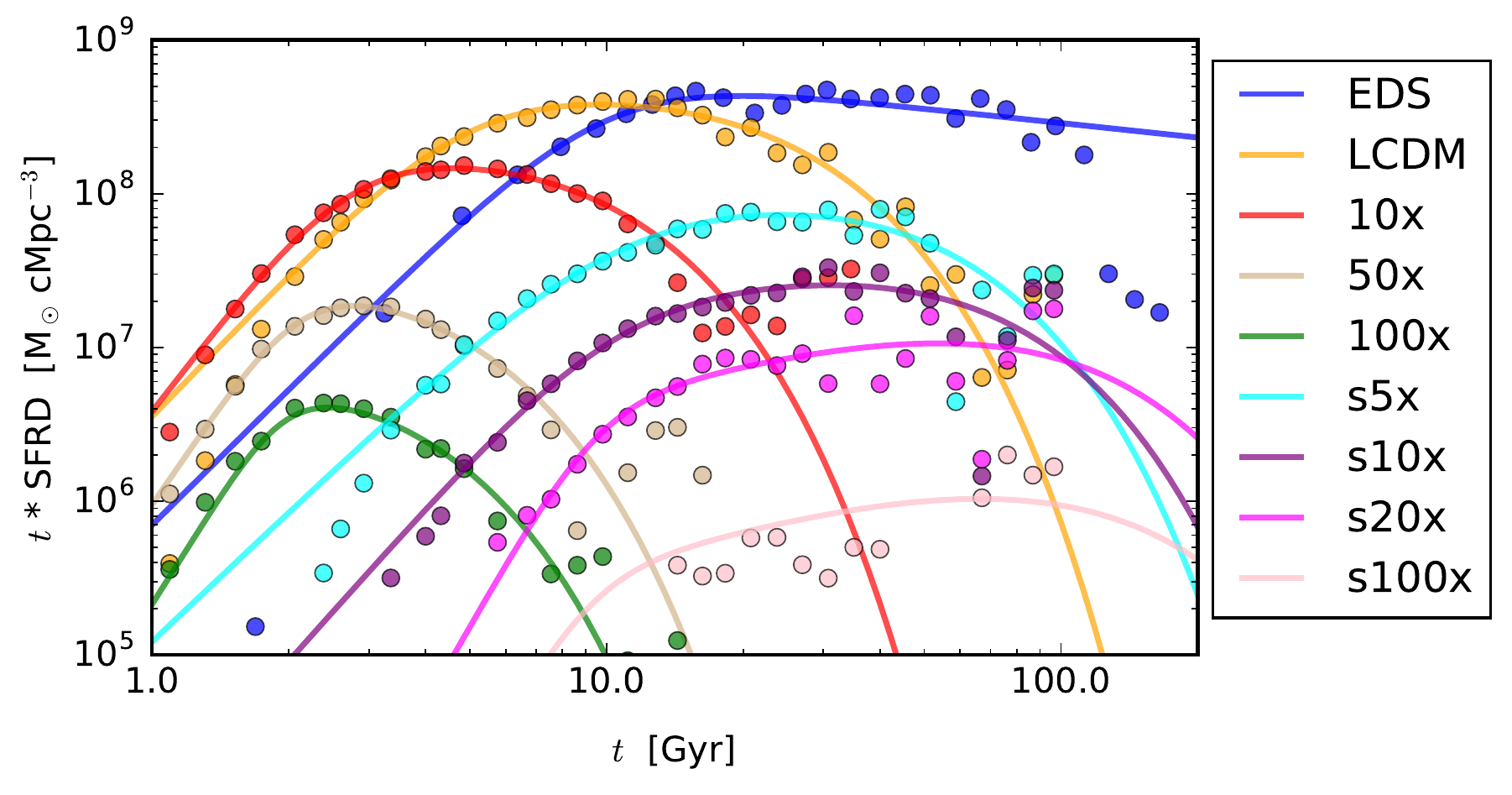}
	}
	\hfill
	\subfloat[Fractional residuals of the SFRD fits. \label{fig:conv-residual}]{%
		\includegraphics[width=.8\linewidth]{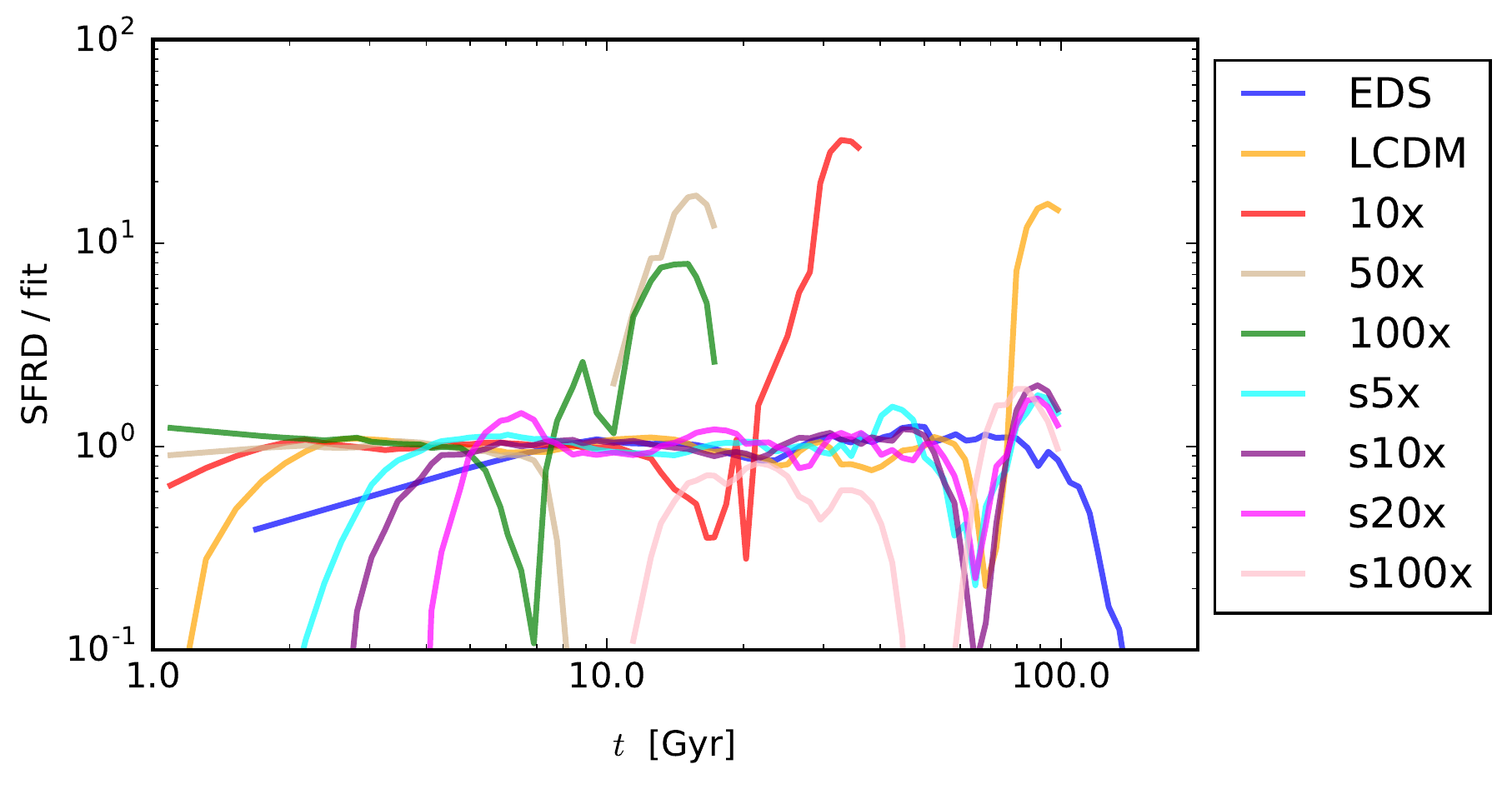}
	}
	\caption{The performance of the double power-law fitting formula (Equation \ref{eq:sfrd_fit}) in describing the SFRD data obtained from our simulations of both counterfactual and {\sl LCDM} universes. The blue, red, tan, green, cyan, purple, magenta, pink and orange lines and dots correspond to {\sl EDS}, {\sl 10x}, {\sl 50x}, {\sl 100x}, {\sl s5x}, {\sl s10x}, {\sl s20x}, {\sl s100x} and {\sl LCDM} respectively; these colours representing different counterfactual universes are consistent with previous plots. In general, the fit successfully captures the evolution of the SFRDs in different universes. However, as shown in the lower plot of residuals, there are significant deviations in fractional precision at late enough times for certain simulations, hinting at issues related to resolution. The reasons for this behaviour are discussed in Section \ref{sec:uvb_scaling}.}
	\label{fig:conv-fitgoodness}
\end{figure*}

For all these star formation histories, it is convenient to have a
smooth fit to the time-dependent SFRD: 
\begin{equation}
{\rm SFRD}(z) = a \frac{(1 + z)^{b}}{1 + \left[(1 + z)/c\right]^{d}}\ \mathrm{M_\odot\ yr^{-1}\ cMpc^{-3}}.
\label{eq:sfrd_fit} 
\end{equation} where $a=0.015, b=2.7, c=2.9, d=5.6$ for the analytic fit to observations \citep{2014ARA&A..52..415M}. It turns out that the SFRD from our simulations can also be fitted accurately with this double power-law in $(1+z)$. We illustrate the goodness of the fit both directly and via its residuals in Figure \ref{fig:conv-fitgoodness}. However, we do see cases where the  computed SFRD rises sharply above the declining fit at late times. This upturn was discussed in detail in \citet{2021MNRAS.507.5432O}, where we established that it marks the maximum time at which the SFRD calculation can be treated as numerically reliable on resolution grounds. The model fits to the SFRD data therefore exclude the data beyond these times.

\begin{table}
	\caption{List of simulations in this paper with their reference name, the values of the best fit parameters of Equation \ref{eq:sfrd_fit} and their corresponding $R^2$ value. The goodness of fit varies significant between the two categories of the counterfactual universes. For universes with scaled $\Lambda$, the fit is much better than that of scaled $\sigma_8$ . Refer to Section \ref{sec:uvb_scaling} for more details.}
	\label{tab:sfrd-fit}
	\begin{tabular}{|p{.12\columnwidth}|p{.12\columnwidth}|p{.12\columnwidth}|p{.12\columnwidth}|p{.12\columnwidth}|p{.1\columnwidth}|}
		\hline
		\multicolumn{6}{|c|}{Simulation setup list} \\
		\hline
		Reference name & $a$ & $b$ & $c$ & $d$ & $R^2$\\
		\hline
		{\sl LCDM} & 0.0259 & 1.56 & 2.35 & 4.70 & 0.983\\
		\hline
		{\sl EDS} & 0.0283 & 1.97 & 1.46 & 4.87 & 0.911\\
		\hline
		{\sl 10x} & 0.0134 & 1.29 & 3.44 & 5.52 & 0.990\\
		\hline
		{\sl 50x} & 0.0222 & 1.31 & 3.72 & 5.89 & 0.992\\
		\hline
		{\sl 100x} & 0.000689 & 1.25 & 3.52 & 5.76 & 0.990\\
		\hline
		{\sl s5x} & 0.00508 & 0.740 & 1.53 & 3.46 & 0.519\\
		\hline
		{\sl s10x} & 0.00148 & 0.540 & 1.48 & 4.03 & 0.551\\
		\hline
		{\sl s20x} & 0.000425 & 0.311 & 1.46 & 6.49 & 0.165\\
		\hline
		{\sl s100x} & 0.000035 & 0.251 & 1.51 & 6.60 & 0.190\\
		\hline
	\end{tabular}
\end{table}

We also note a drop in the {\sl EdS} SFRD above about 100\,Gyr. This drop coincides with the halo mass function ceasing to evolve in our simulation.  This is in contrast with expectations for the {\sl EdS} case, which predicts continual growth of haloes. This lack of evolution reflects our failure to resolve haloes at late times, which eventually limits all our simulations. In the {\sl EdS} case, this appears to bias the star formation to low values; but by this point the total production of stars has converged.

The values of the fitted parameters of Equation \ref{eq:sfrd_fit} are summarised in Table \ref{tab:sfrd-fit}. On top of these parameters, we also include a measure of the precision of the fit, $R^2$: 
\begin{equation}
R^2 = 1 - \frac{\sum_{t=0}^t \left({\rm SFRD}_{\rm sim}(t) - {\rm SFRD}_{\rm fit}(t)\right)^2}{\sum_{t=0}^t \left({\rm SFRD}_{\rm sim}(t) - \overline{\rm SFRD}_{\rm sim}\right)^2}\; ,
\label{eq:sfrd_goodness}
\end{equation}
where $\overline{\rm SFRD}_{\rm sim}$ denotes the mean of the SFRD from the simulation and the rest of the symbols have the same meanings as before. This quantity provides a quantitative complement to the residuals shown in Figure \ref{fig:conv-fitgoodness}. The value of $R^2$ varies between zero and one with higher values corresponding to a better fit.  As we see in Table \ref{tab:sfrd-fit}, the SFRD in counterfactual universes with scaled $\Lambda$ can be fitted in this way to high precision, about as well as our observed universe.


\JAP{
The main potential concern with these results is the impact of the inability of the simulations to follow the halo population into the indefinite future, as discussed above in Section~\ref{sec:evol-hmf-igm}. Provided the critical time at which haloes are lost is beyond the peak in the SFRD, we can have some confidence that the peak in star formation is physical and that the apparent convergence in the total production of stars is robust. The model that comes closest to violating this criterion is {\sl $100x$}, where Figure~\ref{fig:new_cosmo-hmf} shows that haloes are being lost by $t=7$\,Gyr, whereas the peak in the SFRD is at about 2\,Gyr. This particular simulation is the one that has the strongest impact on our understanding of the suppression of asymptotic star-formation efficiency at high $\Lambda$, so we felt it was important to repeat this calculation with a higher resolution than our fiducial choice. As discussed in Section~\ref{sec:evol-hmf-igm}, this remedies the loss of haloes at 7\,Gyr, and so we can consider the associated impact on the SFRD. The runs with two different resolutions are contrasted in Figure~\ref{fig:new_cosmo-sfr}, where it can be seen that the results are very close, both in terms of the peak in SFRD and the subsequent decline. In summary, the haloes are dissolving at a time when the star forming activity is already low. We therefore conclude that our results are a fair representation of the predictions of the {\tt Enzo} model, and are not affected by resolution artefacts. 
}

 

\subsection{UV background scaling}\label{sec:uvb_scaling}

A significant element in the {\tt Enzo} model for star formation is allowance for the effect of a UV background on the thermal evolution of the IGM. This presents a problem of self-consistency, since the level of the UV background depends on the star-formation history that we are aiming to calculate. We can only approach this iteratively: assume a background, compute the SRFD, and then estimate a revised background from this.
We make the simple assumption that the UV background is dominated by young massive stars, whose short lifetime means that their abundance is directly proportional to the SFRD. Therefore, we scale the UV background with the ratio of the SFRDs in each counterfactual universe to {\sl LCDM} at identical times:
\begin{equation} \label{eq:uvb_scaling}
\frac{\Gamma_{\rm scaled}(t)}{\Gamma_{\rm LCDM}(t)} = \frac{{\rm SFRD}_{\rm scaled}(t)}{{\rm SFRD}_{\rm LCDM}(t)},
\end{equation} where $\Gamma$ is the photoheating rate from the UV background, subscripts `scaled' and `LCDM' represent counterfactual and {\sl LCDM} cosmology respectively.
In practice, we implement the UV scaling via the smooth fits to the SFRD discussed above.

We thus ignore any contribution from quasars, which are the most probable alternative source of the UV background. However, the emissivity of quasars appears to be sub-dominant in comparison to galaxies unless there is a significant number of very low luminosity AGN \citep{2005MNRAS.356..596M, 2007MNRAS.382..325B, 2007MNRAS.374..627S}. We assume that this smaller relative contribution applies in all our models. We also neglect more exotic alternative contributions to the UV background, such as an early generation of black holes \citep{2004MNRAS.352..547R, 2005MNRAS.357..207R, 2001ApJ...563....1V}. While possible in principle, there is currently little observational support for significant contributions from these mechanisms \citep{2009RvMP...81.1405M}.

We can now investigate the self-consistency of the SFRDs in counterfactual universes by implementing the scaled UV background. The initial computations assumed the standard UV background from the {\sl LCDM} run, and these were then repeated with the UV background based on this initial SFRD calculation, adopting the scaling from equation \ref{eq:uvb_scaling}. If we refer to Figure \ref{fig:conv-sfrd-fit}, the SFRDs of counterfactual universes relative to {\sl LCDM} suggest that the impact of the scaled UV background on the SFRDs is small. At early times, the SFRDs of all counterfactual universes are similar to {\sl LCDM} or are lower, meaning that deviations in the photoheating rates in these universes will not be as extreme as switching the UV background off completely. At late times, even though the SFRD of the {\sl EDS} cosmology is much higher than {\sl LCDM}, the photoheating rates at these times are already low. Therefore, any effect from a scaled UV background will be less extreme than removing the UV background entirely. In practice, therefore, for computations of the SFRD it seems sufficient to use the \citet{2012ApJ...746..125H} UV background model, even though in principle this should vary with cosmology.

\subsection{IGM with scaled UV background} \label{sec:igm_uvb}

\begin{figure*}
	\centering
	\subfloat[Scaled $\Lambda$ simulations \label{fig:luvb}]{%
		\includegraphics[width=0.6\linewidth]{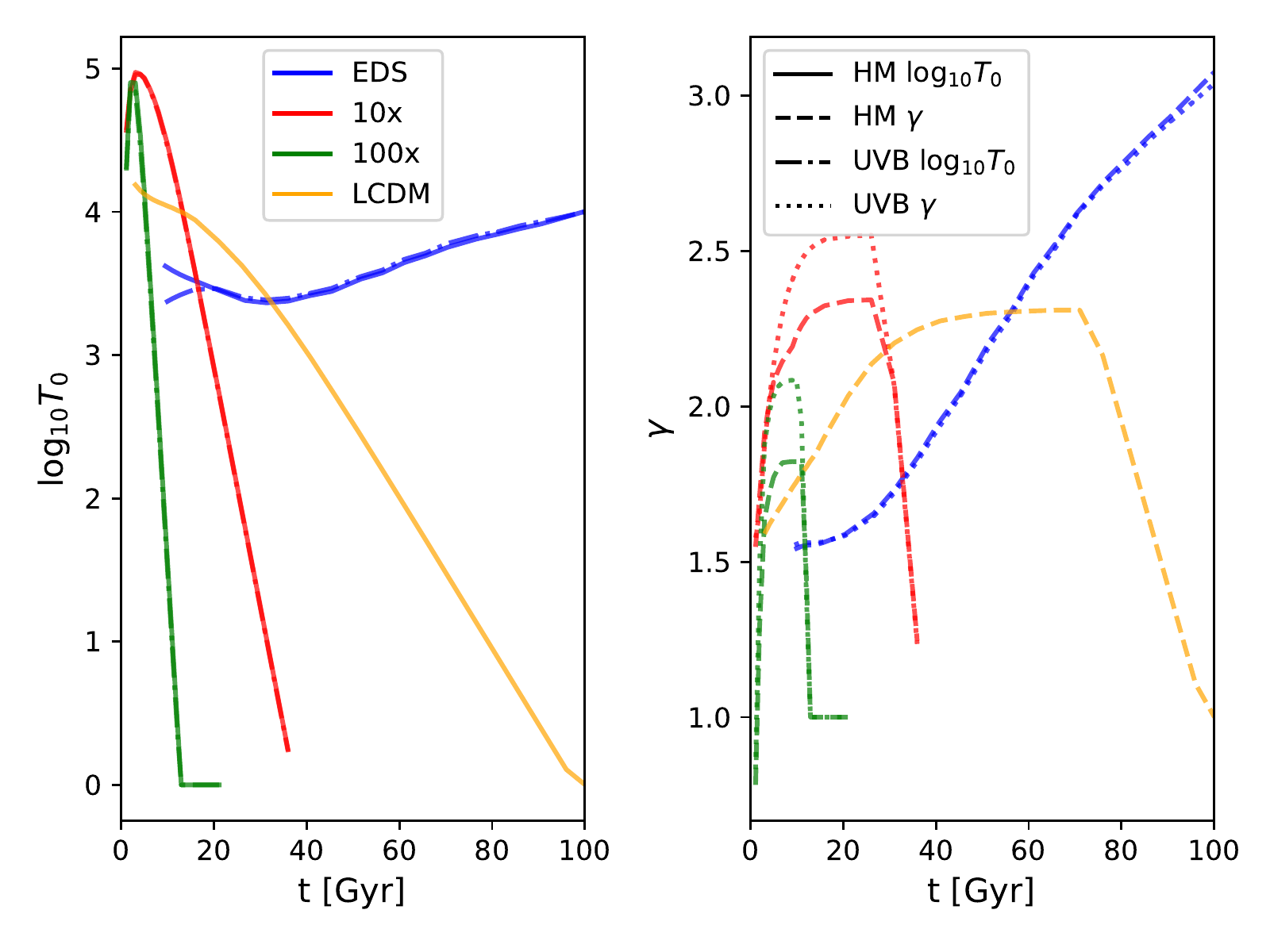}
	}
	\hfill
	\subfloat[Scaled $\sigma_8$ simulations \label{fig:suvb}]{%
		\includegraphics[width=0.6\linewidth]{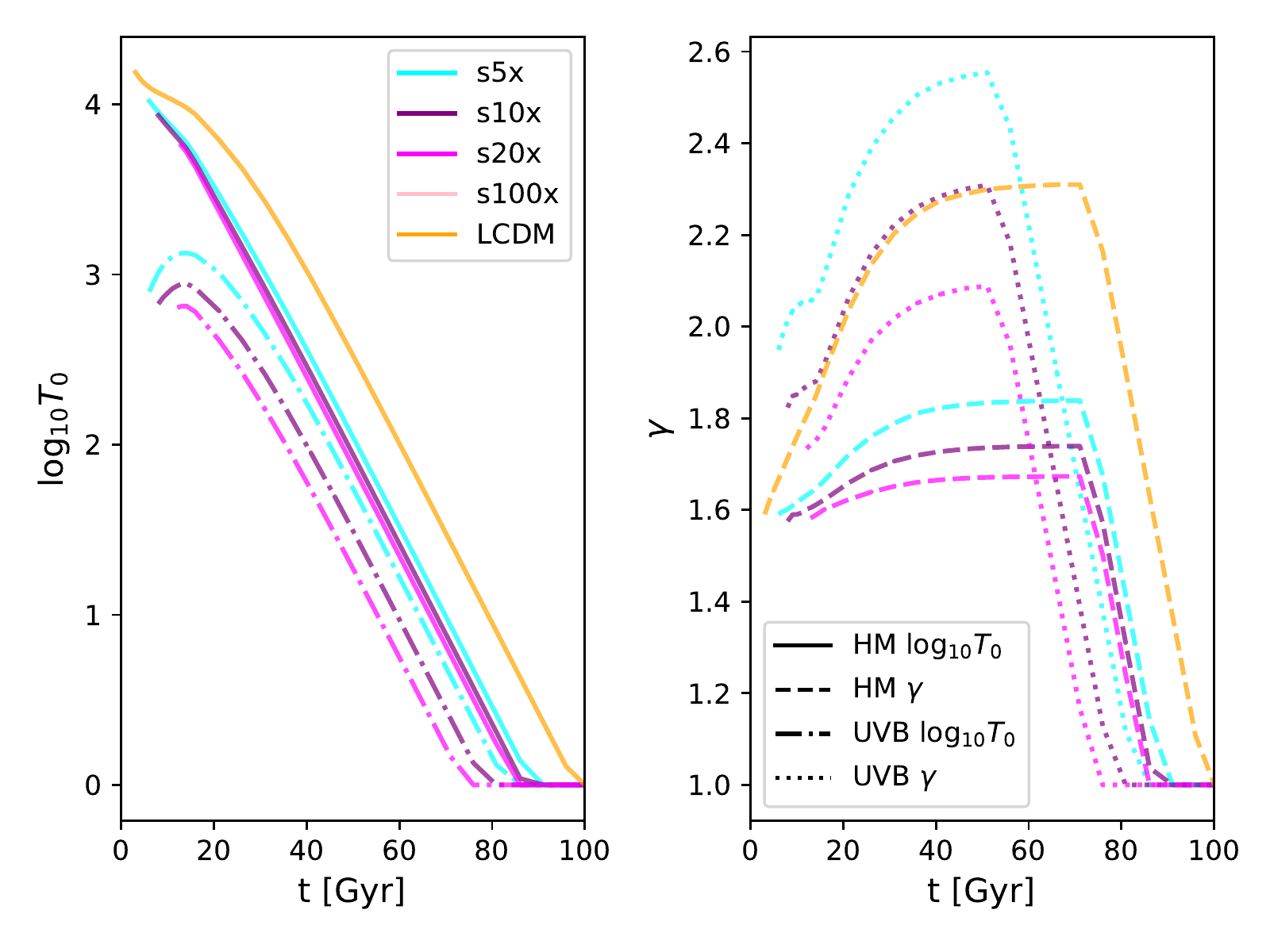}
	}
	\caption{Evolution of $\gamma$ and $T_{\rm 0}$ across cosmic time for simulations with scaled UV background. The lack of data at early times arises because of a lack of gas within the specified overdensity used to calculate $T_0$ and $\gamma$ (see the main text). The various cosmologies are colour coded in a similar way as in Figure \ref{fig:conv-fitgoodness}. The simulations are further categorised into those with the \citet{2012ApJ...746..125H} UV background (HM) and the scaled UV background (UVB); these differences are indicated by varying the line style appropriately for each colour. Note that there is no UVB line in the case of {\sl LCDM}, as this is the fiducial simulation for scaling the HM background. The impact of the scaled UV background is more significant for cosmologies with scaled $\sigma_8$ than for those with scaled $\Lambda$, as may be seen by comparing solid versus dash-dotted lines and dashed versus dotted lines. We discuss these differences in Section \ref{sec:igm_uvb}.
	\label{fig:conv-uvb-igm}}
\end{figure*}

Even though the history of star formation turns out not to have a strong dependence on the assumed UV background, this hides some of the underlying complexity of the situation, and we show in this Section that the properties of the IGM do certainly depend on the assumed level of the UV radiation field.
Following \citet{2021MNRAS.507.5432O}, we focus on the phase distribution of the IGM material, defined as having an overdensity less than $10^3$ \citep{2001ApJ...552..473D}. We adopt the power-law model of \citet{1997MNRAS.292...27H}, in which the IGM temperature is given by
\begin{equation}
\label{eq:igm_eos}
T = T_0(1+\delta)^{\gamma-1}, 
\end{equation} 
where $T_0$ is the temperature at the mean density, $\delta$ is the gas overdensity and $\gamma$ is a sensitivity parameter for the equation of state.
We show the differences in these IGM parameters between simulations with the default \citet{2012ApJ...746..125H} UV background (HM) and the scaled UV background (UVB) in Figure \ref{fig:conv-uvb-igm}, assuming that the SFRD remains exactly the same in each universe regardless of the UV background. We break down the analysis into universes with a scaling of $\Lambda$ (top) and $\sigma_8$ (bottom).

\begin{figure*}
	\centering
	\includegraphics[width=0.6\linewidth]{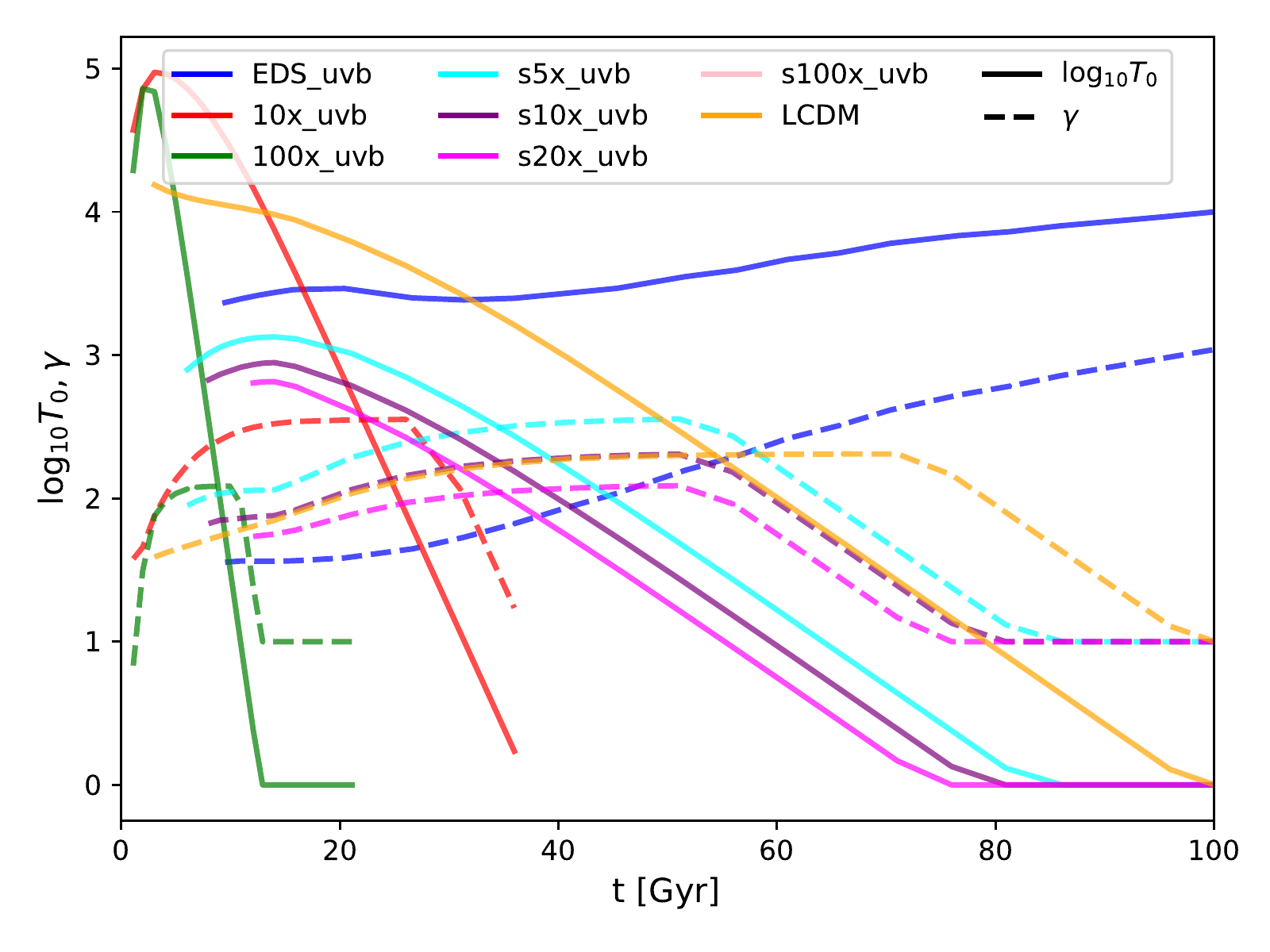}
	\caption{Evolution of $\gamma$ and $T_0$ across cosmic time for simulations with scaled UV background. The various cosmologies are colour coded in a similar way as in Figure \ref{fig:conv-fitgoodness}. The IGM evolves more drastically in a short period of time when we scale $\Lambda$. On the other hand, the IGM in {\sl EDS} becomes continuously hotter beyond $40\,{\rm Gyr}$. We discuss these differences in Section \ref{sec:igm_uvb}.}
	\label{fig:new-conv-igm}
\end{figure*} 

In Figure \ref{fig:luvb}, we see that the scaled UV background does not affect the evolution of $T_0$. The solid lines (HM: standard Haardt--Madau background) and dashed-dotted lines (UVB: scaled background) are indistinguishable from each other. But the equation of state is affected: at intermediate times, the HM $\gamma$ is consistently lower than the UVB $\gamma$, indicating that the temperature of the IGM is more sensitive to its overdensity in the simulations with a scaled UV background. The reduced heating from the scaled UV background makes cooling the dominant process, which is sensitive to density.

On the other hand, the IGM in the {\sl EDS} simulation shows an almost identical evolution regardless of the UV background. The only difference occurs in the value of $T_0$ before
$20\,{\rm Gyr}$, when the SFRD reaches a peak. Before this time, the photoheating rate in the UVB run is lower than the HM case because the SFRD of {\sl EDS} is lower than that of {\sl LCDM}. Therefore, the resulting value of $T_0$ in the UVB case is slightly smaller than for HM, although this slight difference is not sufficient to affect $\gamma$. Beyond this time, the HM photoheating rates are low, reducing the impact of any scaling of the SFRDs, so that the subsequent evolution of $T_0$ and $\gamma$ is independent of the UV background.

In contrast to these results, $T_0$ for the scaled $\sigma_8$ simulations varies significantly according to changes in the UV background, as shown in in Figure \ref{fig:suvb}. $T_0$ values for simulations with a scaled UV background are consistently lower than for their counterparts with an HM UV background. If we compare the SFRDs of {\sl s5x}, {\sl s10x}, {\sl s20x} and {\sl s100x} with {\sl LCDM}, they are significantly lower, translating into a weaker UV background. This reduced heating rate also causes $T_0$ to reach the simulation's assumed temperature floor of $1\,{\rm K}$, which correspondingly causes $\gamma$ to approach a value of unity more rapidly. The opposite is true for {\sl EDS}, for which the SFRD is higher than for {\sl LCDM}. This difference leads to the increase in $\gamma$ at late times seen in Figure \ref{fig:luvb}. Simulations with a scaled UV background thus experience a slightly different evolution from the HM calculation.  We will therefore henceforth consider only the more self-consistent analyses of the IGM that use simulations with a scaled UV background.

With this restriction to a single prescription for the UV background, Figure \ref{fig:new-conv-igm} superimposes the evolution of the IGM associated with all the various counterfactual universes into a single panel. At the start of the simulations, $T_0$ in {\sl 10x} and {\sl 100x} is higher than in {\sl LCDM} for a short period of time. If we compare the SFRDs of these simulations in Figure \ref{fig:conv-sfrd-fit}, we find that the initial SFRDs in {\sl 10x} and {\sl 100x} are slightly higher than in {\sl LCDM}. This difference raises the initial heating rates, resulting in a higher $T_0$. The IGM then cools rapidly to the temperature floor, because of a combination of a much lower SFRD and a faster expansion as compared to {\sl LCDM}. As a result the equation of state also rapidly flattens to $\gamma = 1$. The evolution of the IGM that takes place over a period of approximately $100\ {\rm Gyr}$ in {\sl LCDM} is essentially compressed into $40\ {\rm Gyr}$ for {\sl 10x} and even further into $20\ {\rm Gyr}$ for {\sl 100x}. We also observe a similar trend for simulations with scaled $\sigma_8$. A lower $\sigma_8$ leads to less star formation, resulting in a lower heating rate from the UV background. Therefore, simulations with larger downward scaling of $\sigma_8$ experience a faster decline of $T_0$ towards zero and a similarly rapid flattening of $\gamma$.

In summary, parameter scaling determines the speed at which cosmic evolution causes the characteristic IGM temperature $T_0$ to approach the temperature floor and the equation of state slope $\gamma$ to flatten, but the behaviour is different for universes with scaled $\Lambda$ and $\sigma_8$. For the former, the histories of the IGM in all models follow a common track at high $z$, departing from this at the point where  
$\Lambda$ dominates. In the simulations where $\sigma_8$ is lowered, the initial state is already different, with {\sl LCDM} always having a higher $T_0$ than the {\sl s$\beta$x} simulations. The reduced clustering also leads to both $T_0$ dropping to the temperature floor more rapidly, accompanied by a faster flattening of $\gamma$.

Given these substantial changes to the IGM, it is stiking that there is little to no directly resulting effect on the SFRD. At late times when the universe is dominated by $\Lambda$ and the UV background is insignificant, the gas in the IGM cools to the temperature floor regardless. At early times, we find that the absence of the UV background affects gas of low overdensity most significantly, and this material lies below the threshold for star formation.

\subsection{Average metallicity of young stars} \label{sec:metallicity}

\begin{figure*}
	\centering
	\includegraphics[width=0.6\linewidth]{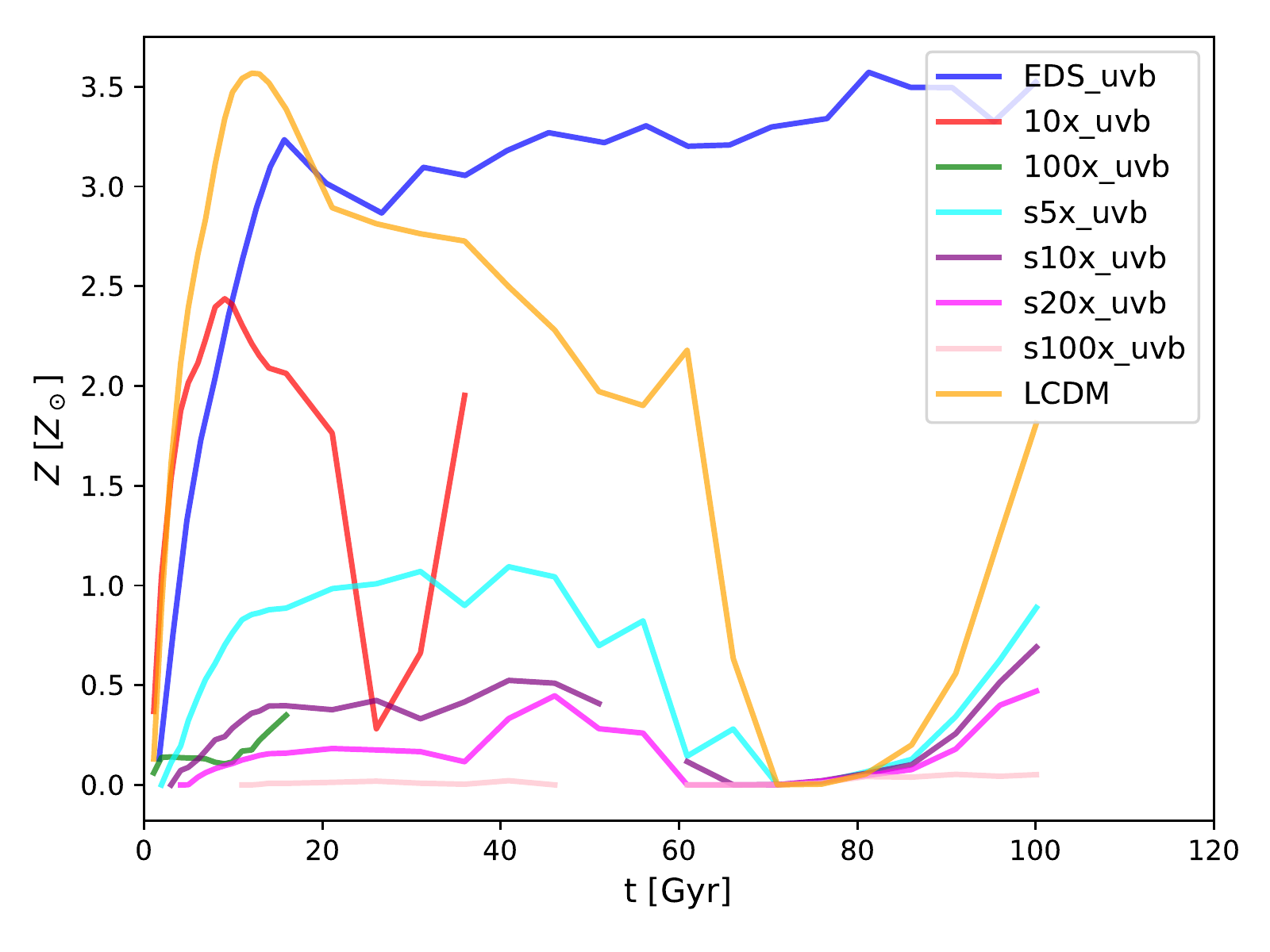}
	\caption{Evolution of the average metallicity of stars formed within $500\ {\rm Myr}$ of a given time. The various cosmologies are colour coded in a similar way as in Figure \ref{fig:conv-fitgoodness}. Sometimes, there are gaps in the line because there are no stars formed within the specified period at that snapshot. We find that the metallicity scales according to the SFRD. Therefore, these curves have a similar evolution to the trends shown in Figure \ref{fig:new_cosmo-sfr}. Refer to the discussion in Section \ref{sec:metallicity}.}
	\label{fig:metallicity}
\end{figure*}

An interesting aspect of the long-term star formation history is its associated chemical evolution. Will stars continue to form from ever more polluted gas in their vicinity, or will there be time for pristine gas from the IGM to mix?  We investigate this by looking at how the average metallicity of young stars ($< 500\ {\rm Myr}$) evolves as a function of time in our counterfactual universes. As described in Section 2 of \citet{2020MNRAS.497.5203O}, {\tt Enzo} tracks metals as part of its feedback prescription, following the injection of metals into the IGM from star-forming activity, and assigning an appropriate metallicity to each newly-formed star particle. We plot the resulting average metallicity as a ratio to the solar metallicity against time in Figure \ref{fig:metallicity}. Except for {\sl EDS} and {\sl s100x}, all other simulations exhibit a similar evolution in the average metallicity of their young stars: this increases up to a peak, declines, and finally starts to increase again at very late times. This final increment in metallicity might plausibly be attributed to the isolated evolution of haloes as a result of freezeout: the haloes could then function as closed boxes, so that the IGM becomes ever more polluted, raising the metallicity of stars formed at late times. However, we note that this feature seems to occur at the same point as the spurious star formation associated with worsening resolution in these models, and we therefore assume that this feature in the metallicity history is similarly not physical.

The general behaviour is then that the stellar metallicity scales with SFRD in the simulation. With the peak metallicity of stars coinciding with the peak of SFRD, the decline in SFRD is also associated with a drop in metallicity. When the simulation is not actively forming stars, there is time for the fuel store of potential future star formation to mix with the inflow of fresh pristine gas from the IGM. Because of the difference in the gas cooling and star formation time scale discussed earlier, the metallicity of the young stars decreases. But in any case, the peak metallicity is lower in the highly scaled models where the total star-forming activity is suppressed, reflecting the lack of metal pollution of the IGM. Conversely, with the active star formation seen in {\sl EDS}, the gas is continuously injected with new metals associated with feedback. Therefore, after the initial rise in metallicity, stars in {\sl EDS} remain highly enriched due to the consistently higher SFR shown in Figure \ref{fig:new_cosmo-sfr}.

\section{Asymptotic star formation efficiency and observer selection}
\label{sec:asymptotic}

In the previous sections, we have surveyed the astrophysical issues that determine the history of star formation in a given counterfactual universe with altered cosmological parameters. We now move on to the broader question of the effect of such variations in SFRD within an {\it ensemble\/} of universes. The motivation is to use anthropic reasoning as a potential means of addressing puzzlingly unnatural values of cosmological parameters -- pre-eminently the small value of the cosmological constant. The philosophy here is as originally set out by \citet{1987PhRvL..59.2607W, 1989RvMP...61....1W}: models in which structure formation is heavily suppressed must contain fewer stars and hence fewer observers. It is simplest to think of this reasoning in the context of a physical multiverse, in which many distinct universes `really' exist (as in, for example, the case of bubble universes in eternal inflation -- see e.g. \citejap{Linde2015}). But if such bubbles are causally disconnected, it is reasonable to wonder what difference their physical existence can possibly make, and in fact the Bayesian analysis of the situation is identical whether or not the other universes exist as concrete entities or merely possibilities -- in the same way that it is possible to assert that the probability of a fair coin landing heads is 0.5, without needing to toss it many times. Concentrating on the case of varying $\Lambda$, we would write
\begin{equation}
P(\Lambda\mid {\rm observer}) \propto
P(\Lambda)\, P({\rm observer}\mid \Lambda),
\end{equation}
where $P(\Lambda)$ is the prior probability density of $\Lambda$, and $P({\rm observer}\mid \Lambda)$ is the observer weighting to be applied. Both of these factors are far from trivial to establish unambiguously. 

Starting with the prior, and thinking specifically of an inflationary multiverse, seeded bubble universes are formally of infinite volume. It is therefore far from clear how different members of any ensemble should be weighted (see e.g. \citejap{2008PhRvD..77f3516G}). This
`measure problem' was evaded for the case of $\Lambda$ by \citet{1987PhRvL..59.2607W, 1989RvMP...61....1W, 2000astro.ph..5265W}, through the argument that $\Lambda=0$ should not be a special value. Therefore, whatever the unknown form of $P(\Lambda)$ in general, it should be reasonable to treat it as constant in a small range of values around zero. Provided observer weighting then strongly disfavours values that depart very far from zero, this argument seems acceptable.

Observer weighting is similarly a thorny issue. We have the intuitive sense that a member of the ensemble with  more observers is more likely to be the one that is experienced, assuming that we randomly select a single observer from the totality of possible observers. This reasoning is successful in disposing of the otherwise disturbing `doomsday paradox' (see e.g. \citejap{Olum_Doomsday}), but remains difficult in detail. For example, are ants cosmological observers? Should long-lived creatures who can observe the universe on many more occasions be given greater weight? These and more subtleties are explored by \citet{Neal2006}. To some extent, such worries can be evaded by arguing that we are only seeking relative probabilities, and these become easier to consider as the multiverse ensemble becomes simpler. If we were studying ensembles where subatomic physics varied, we might need to debate the relative merits of carbon-based and silicon-based life, for example; this is a question best avoided by cosmologists. But in the case where all of physics apart from $\Lambda$ is retained unaltered, we can take a simpler approach. Weinberg's argument for a uniform prior might equally be viewed as giving the probability distribution for the value of $\Lambda$ associated with a single baryon randomly chosen from the multiverse ensemble. In order to apply observer selection, then we are led to ask ourselves about the efficiency of turning baryons into intelligent observers. Making the minimal assumption that the complex structures associated with star formation are needed for the generation of observers, we can see that a natural candidate for the observer weighting factor is the global efficiency of star formation, and we adopt this in what follows.

In early work on this topic, the required efficiency was estimated using a simple collapse argument, first set out in detail by \citet{1995MNRAS.274L..73E}. Here one takes the empirical fact that most stars exist in galaxies of about the mass of the Milky Way. Thus there is a critical mass of dark-matter halo, in the region of $10^{12}\,{\rm M}_\odot$, and the interesting question is the fraction of mass in the universe that has undergone gravitational collapse into such haloes, with the presumed average efficiency of conversion of baryons into stars being proportional to this collapse fraction. This fraction is small at early times, but grows as gravitational instability progresses, until it freezes out at some asymptotic value as $\Lambda$ comes to dominate. This single-scale model actually accounts very well for the observed evolution of the cosmic stellar
density from high redshift to the present \citep{2007MNRAS.379.1067P}. In this paper, we will therefore look at the effect of replacing this simple analytic estimate of the star-formation efficiency with one taken directly from simulation. Note that we are interested in the asymptotic efficiency: the fraction of baryons that are processed through stars, independent of when this happens. The assumption is that there is nothing special about our current era: observers could have existed at high redshift, and will also be associated with the stars that form in the very distant future.

\begin{figure*}
	\centering
	\includegraphics[width=0.65\linewidth]{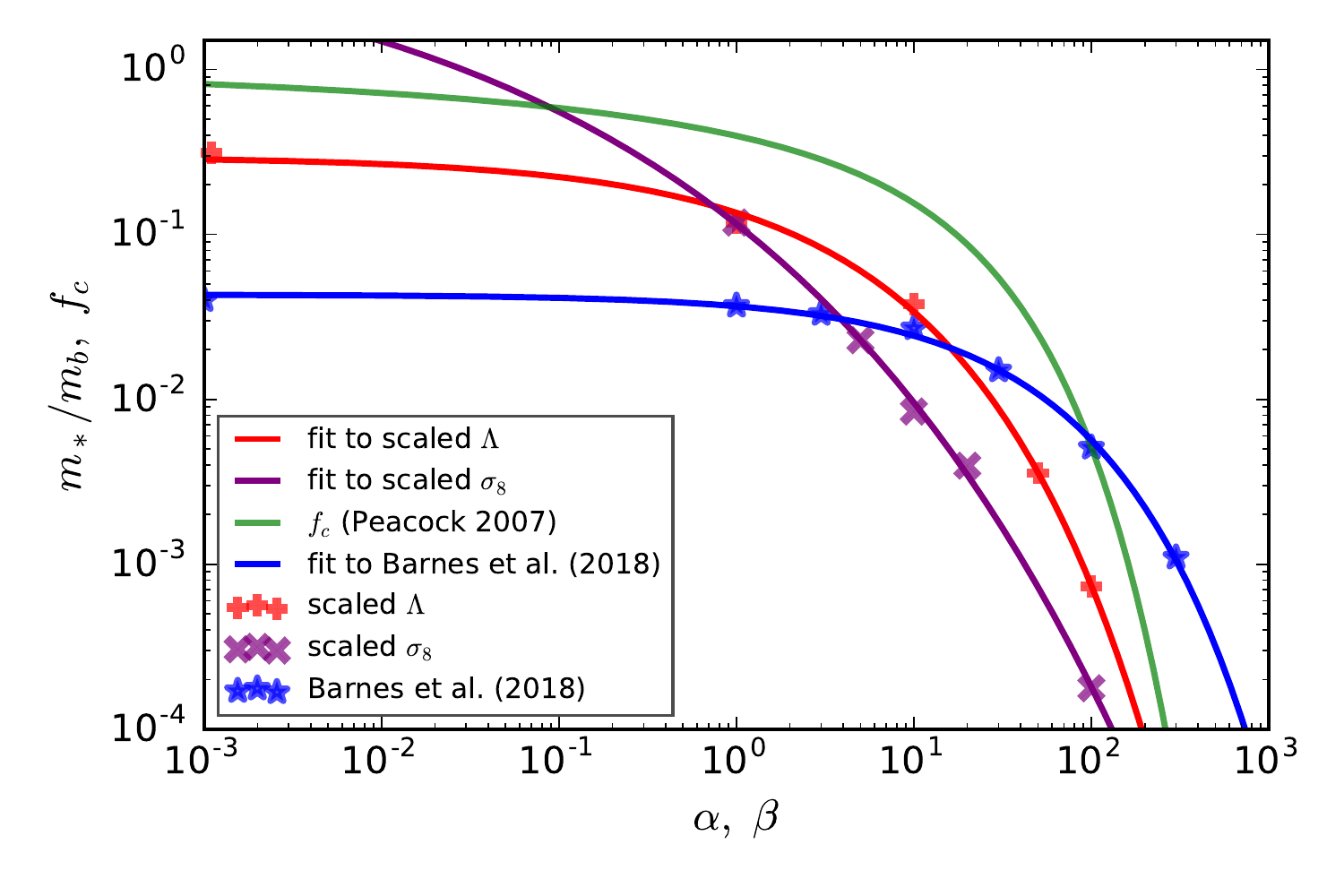}
	\caption{This plot shows the observer weighting from the asymptotic efficiency of star formation, estimated either using the collapse fraction from \citet{2007MNRAS.379.1067P} or directly from simulations as a function of the scaling factor for $\Lambda$, $\alpha$, and the scaling factor for $\sigma_8$, $1/\beta$. 	We include results for the {\sl EDS} model by plotting them at $\alpha=0.001$ rather than the true $\alpha=0$. The green line is the collapse fraction, with the red pluses and purple crosses representing respectively the asymptotic star formation efficiency from our scaled $\Lambda$ and scaled $\sigma_8$ simulations. The blue points are taken from \JAP{Figure 5 of} the EAGLE simulations by \citet{2018MNRAS.477.3727B}. We also show fits to the various simulation results with lines of the same colour as the data points. The fits to $f_c$  and $m_* / m_b$ from our scaled $\Lambda$ and scaled $\sigma_8$ simulations are given by Equations \ref{eq:fc_fit}, \ref{eq:asfl_fit} and \ref{eq:asfs_fit} respectively, and the fit to \citet{2018MNRAS.477.3727B} by Equation \ref{eq:asfl_barnes}.
	}
	\label{fig:anthropic_fit}
\end{figure*}

In order to estimate this asymptotic production of stars, we refer to the fits obtained for the counterfactual universes in Table \ref{tab:sfrd-fit}. Since the simulations are all terminated at some finite time, we can only obtain the total production of stellar mass by taking the fits to the SFRD that we found in Equation \ref{eq:sfrd_fit} and extrapolating these to infinite time:
\begin{equation}
m_* (t \to\infty) = V \int_{0}^{\infty} {\rm SFRD}(z)\ dt,
\end{equation} where $V$ is comoving volume of the simulation. Through the conversion
\begin{equation}
\frac{dz}{dt} = -\frac{\dot{a}}{a^2} = -\frac{H(z)}{a},
\end{equation} we can then obtain the asymptotic total stellar mass produced by the simulation
\begin{equation}
m_* (t \to\infty) = V \int_{-1}^{\infty} \frac{{\rm SFRD}(z)}{H(z)(1+z)}\ dz,
\end{equation} where the limits reflect the infinite lifetime of the universes. The total available baryon mass is
\begin{equation}
m_b = V \times 2.7755\e{11}\Omega_b h^2\; M_\odot,
\end{equation} where the cosmological parameters are as appropriate for each counterfactual universe. In practice, the way we have altered $V$ with the scaling of the cosmological parameters means that this mass takes the same value of $10^{16.54}\,M_\odot$ for all models (See Section \ref{sec:ic-scaling}). We then plot $m_*/m_b$ against the $\Lambda$ scaling factor, $\alpha$, in Figure \ref{fig:anthropic_fit}. The specific values of $m_*/m_b$ for our simulations with $\alpha=(0,1,10,50,100)$ are
\begin{equation}
    m_*/m_b = (0.31,\; 0.12, \; 0.038, \; 0.0036, \; 0.00073).
\end{equation}


Figure \ref{fig:anthropic_fit} also compares the dependence on $\alpha$ of this directly computed figure of merit, $m_*/m_b$, with the simple expectation of a single-scale collapse fraction, $f_c$ (assuming a critical halo mass of $10^{12.4}\,M_\odot$, following \citejap{2007MNRAS.379.1067P}). The single-scale curve displays an exponential decline as a function of $\alpha$, the scaling factor for $\Lambda$, and this can conveniently be described with high accuracy by the fitting formula
\begin{equation}\label{eq:fc_fit}
f_c(\alpha) =
 \exp\left(-0.086\,\alpha^{0.78} - 0.84\,\alpha^{0.21}\right).
\end{equation}
This same formula as a function of $\beta$ also yields the asymptotic collapse factor for models with scaled $\sigma_8$. This is by design: $\sigma_8$ is suppressed by a factor $\beta^{1/3}$, so that $\beta\simeq \alpha$ should yield the same frozen-out $f_c$ as a model in which $\Lambda$ is scaled up by a factor $\alpha$.

We have also used a similar analytic form to describe the simulation results for $m_*/m_b$, and how they vary with scaled $\Lambda$. Fewer parameters are used, reflecting the limited number of simulation results:
\begin{equation}\label{eq:asfl_fit}
\frac{m_*}{m_b}(\alpha) = 0.29 \exp\left(-0.78\,\alpha^{0.44}\right).
\end{equation} 
We can also describe the results from
\citet{2018MNRAS.477.3727B} with an equation of the same form:
\begin{equation}\label{eq:asfl_barnes}
\frac{m_*}{m_b}(\alpha) =0.043\, \exp\left(-0.16\,\alpha^{0.55}\right).
\end{equation}
In both cases, the absolute normalization is unimportant; in an ensemble approach, the posterior for $\alpha$, $p(\alpha)$ is proportional to $m_*/m_b$, normalized by integration over $0<\alpha<\infty$.

For completeness, we can use the same analytic form to approximate the asymptotic star formation efficiency from simulations with scaled $\sigma_8$:
\begin{equation}\label{eq:asfs_fit}
\frac{m_*}{m_b}(\beta) \propto \exp\left(-4.2\,\beta^{0.20}\right),
\end{equation} 
where $1/\beta^{1/3}$ is the scaling factor for the normalization. This equation differs from the dependence on $\alpha$, even though simple considerations of collapse fraction would suggest that the dependence should be identical, as discussed earlier.

Concentrating now on the case of scaled $\Lambda$, it is interesting to note the similarities and differences between our simulation results and the single-scale model. Both approaches predict a substantial suppression of the asymptotic star-formation efficiency with increasing $\Lambda$: when $\Lambda$ is scaled up by a factor $\alpha=100$, the simulation results predict a fall in efficiency by close to a factor 100, in comparison with a fall by a factor 30 in the single-scale model. The single-scale values are however larger at all values of $\alpha$: by a factor 3 at $\alpha=1$ and a factor 10 at $\alpha=100$, so that the predicted decline with increasing $\Lambda$ is less marked. In part, this difference could be removed by choosing a different critical mass, and the simulation results are best matched by choosing $10^{13.0}\,M_\odot$ -- although this choice would give a less good match to the observed stellar density as a function of redshift.

With these fits to the star-formation efficiency in hand, we can now calculate the posterior distribution of $\Lambda$, assuming a uniform prior in $\Lambda$. Normalizing by integration between $\Lambda=0$ and $\infty$, we thus obtain a probability distribution for $\Lambda$, subject to the assumption that $\Lambda>0$. As discussed above, a rather different calculation would be required in order to estimate the impact of observer selection in the case of a negative cosmological constant. There, we would expect star formation to be highly efficient in the high-density late stages of recollapse, so the key question is whether these stars form so close in time to the big crunch singularity that there is no chance for associated observers to develop. The observed positive $\Lambda$ gives us a hint that the posterior weight of $\Lambda<0$ is truncated in this way, although it would obviously be important if we were able to demonstrate this from first principles. For the present, we can only proceed on the assumption that the total posterior probability that $\Lambda>0$ is not small, so that we make no important error in normalizing the probability to unity in this regime.

\begin{figure*}
	\centering
	\includegraphics[width=0.65\linewidth]{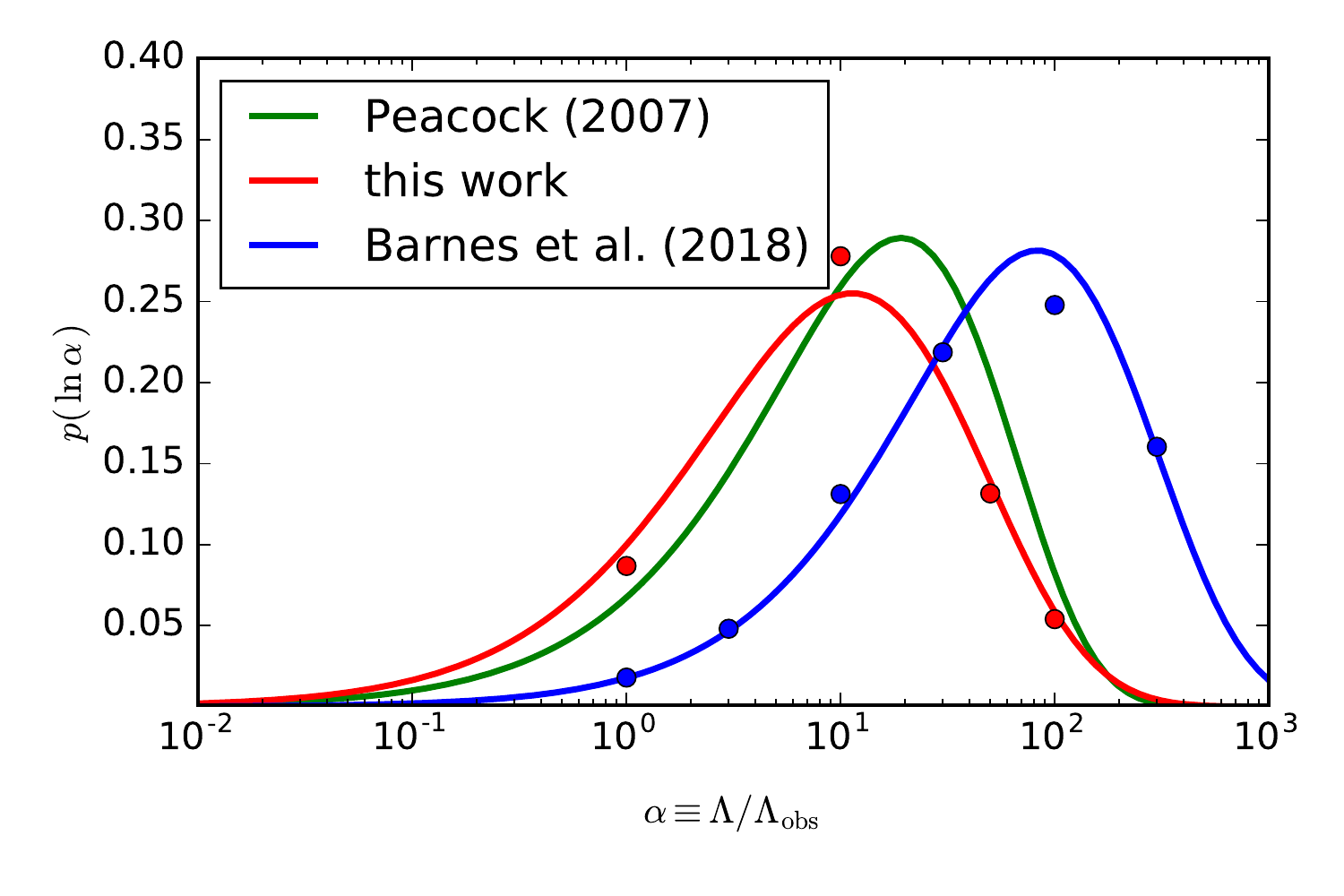}
	\caption{Relative probability per unit $\ln\alpha$ from this work, \citet{2007MNRAS.379.1067P} and \citet{2018MNRAS.477.3727B}. The green, red and blue lines are respectively the fits to the collapse fraction from \citet{2007MNRAS.379.1067P} with Equation \ref{eq:fc_fit}, $m_* / m_b$ from this work and that of \citet{2018MNRAS.477.3727B}, using respectively Equations \ref{eq:asfl_fit} and \ref{eq:asfl_barnes}. The blue dots are extracted from the left panel of Figure 12 from \citet{2018MNRAS.477.3727B}. The results from \citet{2018MNRAS.477.3727B} assign a higher weight to universes with larger value of $\Lambda$ than our work. Both numerical calculations agree with the analytic estimate that the observed value of $\Lambda$ is smaller than the typical value experienced over an ensemble. But our results indicate that the inconsistency is not marked, with $\sim 10\%$ of observers experiencing $\Lambda$ no larger than the observed values. In contrast, \citet{2018MNRAS.477.3727B} estimate this probability to be only $\sim 1\%$, and the two numerical studies thus reach interestingly different conclusions concerning the viability of the simplest anthropic explanation for the observed value of $\Lambda$.
	}
	\label{fig:anthropic_compare}
\end{figure*}

The resulting posterior for $\Lambda$ is displayed in Figure 
\ref{fig:anthropic_compare}. We see a clear truncation of the distribution at large $\alpha$, which can be quantified in a number of ways. One approach would be to quote the upper limit at a certain confidence, or perhaps a confidence range (since $\alpha$ sufficiently close to zero would be considered surprising. Alternatively, we can quote the probability that $\alpha<1$: since the essence of the cosmological constant problem is that natural values are much larger than observed,
the fraction of observers residing in universes whose vacuum densities are no larger than ours gives a measure of how well observer selection succeeds in solving the problem. These various probabilities are collected in Table~\ref{tab:conflims}.

\begin{table}
	\caption{Posterior confidence limits on $\alpha\equiv\Lambda/\Lambda_{\rm obs}$ according to either our simulated results or the single-scale approximation, based on integration of the fitting formulae (\ref{eq:fc_fit}) \& (\ref{eq:asfl_fit}).
	}
	\label{tab:conflims}
	\centering
	\begin{tabular}{|p{.2\columnwidth}|p{.25\columnwidth}|p{.25\columnwidth}|}
		\hline
		& Single-scale & Simulation\\
		\hline
		$p(\alpha<1)$ & 0.082 & 0.129 \\
		95\% limit & $\alpha<82$ & $\alpha<74$ \\
		99\% limit & $\alpha<145$ & $\alpha<154$ \\
		Median & $\alpha=12.6$ & $\alpha=8.0$ \\
		95\% range &
		$0.56 < \alpha < 110$ &
		$0.32 < \alpha < 105$ \\
		\hline
	\end{tabular}
\end{table} 

These anthropic predictions are in slight tension with our single datum: an observed universe with $\alpha=1$. The median value over an ensemble of observers is roughly an order of magnitude larger than observed, and only about 10\% of observers would experience a cosmological constant as large as ours or smaller. But such a disagreement has little statistical strength given the breadth of the distribution. The 95\% confidence range spans a factor 100 in $\Lambda$, and the observed value lies within this range, which is all that we can reasonably demand from a viable theory. We may take some encouragement from the fact that the simulated results tend to be smaller by some tens of percent than the simple single-scale estimate, increasing the weight of universes like our own.
We can also note that
these predictions will be be affected to some extent by allowance for metal-weighting, which reduces the weights of large $\alpha$ counterfactual universes because the reduced SFRD leads to fewer metals for planet formation (see Figure \ref{fig:metallicity}).  According to the results of \citet{2018MNRAS.477.3727B}, this will decrease the typical values of $\Lambda$ by approximately a factor 2 (see their Figure 12).

However, our simulation results
already favour somewhat lower values of $\Lambda$ than those of \citet{2018MNRAS.477.3727B}. Using the same posterior that weights by stellar mass, their  predicted median value of $\Lambda$ was 50 to 60 times larger than the observed value. Also, they determined that only about 2\% of observers should reside in a universe with a value of $\Lambda$ equal to or less than ours. \citet{2018MNRAS.477.3727B} therefore concluded that the impact of $\Lambda$ on structure formation does not straightforwardly explain the small observed value of $\Lambda$. As we have seen, our simulations do not find such a high level of disagreement. But rather than focusing on the differences between the outcomes of the calculations, we should actually find it remarkable that two independent codes, using rather different numerical approaches and implementing rather different subgrid prescriptions for star formation, should make such similar predictions for the far future of star formation in universes that are very far from the standard $\Lambda$CDM cosmology. Skeptics of galaxy formation modelling have been heard to argue that such models are fine-tuned to match a variety of observational constrains and so do not qualify as truly predictive theories. But here we see that the predictions of the models have a degree of robustness even when asked to calculate in regimes that are extremely far from their comfort zone. There is qualitative agreement that large values of $\Lambda$ do suppress the asymptotic efficiency of star formation, although the effects of the observed $\Lambda$ are only minor. Thus our observed universe is on the low side of the range of values predicted in a multiverse ensemble. The key question is just how rare a fluctuation this represents, and we have seen that direct galaxy formation codes can differ substantially over this question. The same is true with semianalytic approaches, with \citet{2017MNRAS.464.1563S} obtaining a probability for $\Lambda<\Lambda_{\rm obs}$ of order 10\%, rather similar to our figure, while \citet{2022arXiv220407509S} gives a much more pessimistic estimate.  This theoretical uncertainty in effect broadens the posterior distribution for $\Lambda$, and conclusions regarding the viability or otherwise of the anthropic approach to $\Lambda$ need to take this into account.

\section{Summary and conclusions} \label{sec:new-cosmo-conclusion}

We have presented the first suite of simulations in which {\tt Enzo} has been used to simulate the history of cosmic star formation in counterfactual universes. The initial conditions of the simulations were set by scaling the value of $\Lambda$ at very high redshift, maintaining the cosmology at that time otherwise unchanged, and then integrating forward in time to see how this adjustment affects the subsequent history of star formation. In most cases, our models are followed until well past the point at which $\Lambda$ dominates, so that development of structure freezes out; the exception is the $\Lambda=0$ Einstein--de Sitter model, which we follow to an age of 100\,Gyr. We quantify these counterfactual models in terms of modified parameters within the $\Lambda$CDM framework, and we have shown analytically how to determine these modified parameters at $z=0$ (defined to correspond to an observed CMB temperature of 2.725\,K).
We also performed additional simulations in which $\sigma_8$ is scaled, in order to mimic the main effect from scaling $\Lambda$, which is a reduced amplitude of fluctuations after freezeout.  

We then use the prescription for galaxy formation within {\tt Enzo} established by \citet{2020MNRAS.497.5203O, 2021MNRAS.507.5432O} to analyse the impact of cosmology on the evolution of a range of properties including stellar masses, star formation densities (SFRDs), halo mass functions (HMFs), thermal properties of the intergalactic medium (IGM) and properties of gas and stars. In particular, we fit a double power-law  to the SFRDs of these counterfactual universes, following \citet{2014ARA&A..52..415M}. By integrating the fit to infinite time, we obtain the asymptotic star formation efficiency in these universes, which can be used as a proxy for observer weighting within an anthropic approach to the value of $\Lambda$. We summarise our findings as follows:

\begin{itemize}
	\item Starting from the cosmological parameters of our universe obtained from WMAP-9, we scaled $\Lambda$ and $\sigma_8$ to obtain a total of seven counterfactual universes, maintaining spatial flatness. The initial conditions are then generated using {\tt MUSIC} and evolved with {\tt Enzo} as far as $t \approx 100\,{\rm Gyr}$ whenever possible. At early times, when $\Lambda$ is not dominant, the evolution of the universes with scaled $\Lambda$ should be identical, and our results satisfied this test. 
	
	\item The HMF of different universes is extremely sensitive to the scaling factor applied to $\Lambda$. A higher value of $\Lambda$ becomes dominant at an earlier time, leading to freezeout in the clustering evolution of the dark matter and an HMF that becomes fixed at an asymptotic form.  We see this effect clearly in the evolution with time of the simulated HMFs, where a larger value of $\Lambda$ reduces the mass of the most massive halo that ever forms. Beyond a certain point, however, the evolution of the HMF becomes unphysical and the number of haloes found decreases with time in all mass ranges. This effect arises through resolution effects: isolated bound haloes will have a comoving virial radius that tends to zero at late times, and eventually this shrinkage cannot be followed (see Section \ref{sec:new_cosmo-hmf}). 
	
	\item The star formation histories of the counterfactual universes are vastly different. A peak in SFRD($t$) is found even for the $\Lambda=0$ Einstein--de Sitter case, but the location of this peak is influenced by the value of $\Lambda$. Since the cosmic SFRD differs significantly between models, we took some care to incorporate a self-consistent UV background, scaling this   according to the ratio of the model SFRD to the fiducial $\Lambda$CDM SFRD.
	
	\item In practice, the scaled UV background turn out not to have an important effect on the SFRD. However, it does influence the evolution of the IGM. The temperature of the IGM decreases to the temperature floor more rapidly in the simulations, as a result of the reduced heating from the scaled UV background. At the same time, the temperature of the gas also loses sensitivity to its density at a similar accelerated pace. 
	
	\item We find in all cases that the average metallicity of young stars in the simulations scales with the SFRD. When the simulation is actively forming stars, the metallicity of the newly formed stars is high, as the fuel for star formation is constantly enriched by feedback from the previously formed stars. But once the SFRD declines, this reservoir of gas is then allowed to mix with the primordial gas entering the halo, reducing the average metallicity of the young stars at later times.
	This trend is less marked in the Einstein--de Sitter case, since the SFRD is not so abruptly truncated by $\Lambda$-induced freezeout.
	
	\item Lastly, we employ these results in the anthropic approach to the value of $\Lambda$. We take the prior on $\Lambda$ to be uniform in a small range around zero, following \citet{1989RvMP...61....1W}. This assumption allows the counterfactual universes to be weighted by their asymptotic star formation efficiency, which can then be used as a proxy for the posterior probability distribution for the value of $\Lambda$ experienced by a random observer within a multiverse ensemble. We provide a fit for this function, and estimate that 95\% of observers would experience a value of $\Lambda$ less than 74 times the measured value, and that 13\% of observers would experience a value of $\Lambda$ smaller than the one that we observe. 
\end{itemize}

The results of this paper are thus relatively encouraging as regards the anthropic approach to the value of the cosmological constant. If we are willing to assert that the probability of the existence of observers scales with the number of stars formed in the universe, then high values of $\Lambda$ are exponentially unlikely to be experienced. This suppression has been estimated in the past by simple collapse-fraction arguments, assuming a critical galaxy scale to dominate the production of stars (e.g. \citejap{1995MNRAS.274L..73E}, \citejap{Garriga1999}, \citejap{2007MNRAS.379.1067P}), and our results are not hugely different to these estimates. In both cases, the typical values of $\Lambda$ are expected to be perhaps an order of magnitude larger than what is observed, but the measured value is not a particularly unlikely fluctuation -- especially in our simulations, which prefer slightly smaller values than the simple single-scale estimate. Our simulations also prefer somewhat smaller values than the similar calculations of \citet{2018MNRAS.477.3727B}, raising the probability that $\Lambda<\Lambda_{\rm obs}$ from their 2\% to about 13\%. The smaller figure would certainly lead to some discomfort with the anthropic approach, so it is rather important to carry out further simulations of this sort in order to see which figure is to be preferred. For the present, we should simply remember that there is a significant uncertainty in the theoretical predictions, and this should be folded in to any statistical assessment of the anthropic approach.

Finally, we must remember that
the case of a negative cosmological constant has not been considered in this study. The central idea of the anthropic approach to $\Lambda$ is to argue that $\Lambda=0$ is not a special value, and that there must therefore be a uniform prior covering values either side of zero. Our results can only be taken to support the anthropic approach if we believe that the posterior distribution is dominated by positive values of $\Lambda$, but at present we have no grounds for believing this to be true. It is entirely possible that a careful study of the full case would conclude that most of the observer weighting should be associated with $\Lambda<0$, in which case the observed positivity of the cosmological constant would be inexplicable. There is therefore strong motivation for repeating the modelling in this paper for the case of recollapsing universes. We can be confident that current codes will need substantial modification in order to cope with new and unfamiliar astrophysical regimes, such as star formation within massive haloes in the presence of the rapidly increasing CMB temperatures to be expected at late times. But there can be no doubt over the importance of attempting such calculations.

\section*{Acknowledgements}

BKO and JAP were supported by the European Research Council under grant number 670193. BKO would like to thank Jose O{\~n}orbe and the TMOX group at the Royal Observatory, Edinburgh for many insightful discussions. For the purpose of open access, the author has applied a Creative Commons Attribution (CC BY) licence to any Author Accepted Manuscript version arising from this submission.

\section*{Data Availability}

The data underlying this article will be shared on reasonable request to the corresponding author.




\bibliographystyle{mnras}
\bibliography{allrefs} 



\appendix


\bsp	
\label{lastpage}
\end{document}